\documentclass{aa}
\usepackage{graphicx}
 
\newcommand{\HI}{\mbox{H{\sc i}}}

\begin{document}      

   \title{Ram pressure stripping of the multiphase ISM and star formation in the Virgo spiral galaxy NGC~4330}

   \author{B.~Vollmer\inst{1}, M.~Soida\inst{2}, J.~Braine\inst{3} 
     A.~Abramson\inst{4}, R.~Beck\inst{5}, A.~Chung\inst{6}, H.H.~Crowl\inst{7}, J.D.P.~Kenney\inst{4}, \& J.H.~van~Gorkom\inst{7}}

   \offprints{B.~Vollmer, e-mail: Bernd.Vollmer@astro.unistra.fr}

   \institute{CDS, Observatoire astronomique de Strasbourg, 11, rue de l'universit\'e,
	      67000 Strasbourg, France \and
	      Astronomical Observatory, Jagiellonian University,
	      Krak\'ow, Poland \and
	      Laboratoire d'Astrophysique de Bordeaux, Universit\'e de Bordeaux, OASU, CNRS/INSU, 
	      33271 Floirac, France \and
	      Yale University Astronomy Department, P.O. Box 208101, New Haven, CT 06520-8101, USA \and
	      Max-Planck-Insitut f\"{u}r Radioastronomie, Auf dem H\"{u}gel 69, 53121 Bonn, Germany \and
	      Department of Astronomy and Space Science, Yonsei University, Republic of Korea \and
	      Department of Astronomy, Columbia University, 538 West 120th Street, New York, 
	      NY 10027, USA
              }

   \date{Received / Accepted}

   \authorrunning{Vollmer et al.}
   \titlerunning{Ram pressure stripping and star formation in NGC~4330}

\abstract{
It has been shown that the Virgo spiral galaxy NGC~4330 shows signs of ongoing ram pressure stripping
in multiple wavelengths: at the leading edge of the interaction, the H$\alpha$ and dust extinction curve sharply out of 
the disk; on the trailing side, a long H$\alpha$/UV tail has been found which is located upwind of a long H{\sc i} tail. 
We complete the multiwavelength study with IRAM 30m HERA CO(2-1) and VLA 6~cm radio 
continuum observations of NGC~4330. The data are interpreted with the help of a dynamical model
including ram pressure and, for the first time, star formation. Our best-fit model reproduces qualitatively the
observed projected position, radial velocity of the galaxy, the molecular and atomic gas distribution and velocity field, and the
UV distribution in the region where a gas tail is present. 
However, the observed red UV color on the windward side is currently not reproduced by the model. Based on our model, the 
galaxy moves to the north and still approaches the cluster center with the closest approach occurring in $\sim 100$~Myr.
In contrast to other Virgo spiral galaxies  affected by ram pressure stripping, NGC~4330 does not
show an asymmetric ridge of polarized radio continuum emission. We suggest that this is due
to the relatively slow compression of the ISM and the particular projection of NGC~4330. 
The observed offset between the H{\sc i} and UV tails is well reproduced by our model. Since collapsing and
starforming gas clouds decouple from the ram pressure wind, the UV-emitting young stars
have the angular momentum of the gas at the time of their creation. On the other hand,
the gas is constantly pushed by ram pressure.
We provide stellar age distributions within 3 radial bins in the galactic disk ($R > 5$~kpc).
Deep optical spectra could be used to test the quenching times suggested by our analysis.
The reaction (phase change, star formation) of the multiphase ISM (molecular, atomic, ionized) to ram pressure is discussed 
in the framework of our dynamical model.
}

\maketitle

\section{Introduction \label{sec:intro}}

The environment has a strong influence on spiral galaxies evolving in a cluster.
The ideal place to study the evolution of cluster spiral galaxies is the Virgo cluster,
because it represents the only cluster in the northern hemisphere where one can observe the ISM distribution and 
kinematics of cluster galaxies at a kpc resolution (1~kpc $\sim 12''$)\footnote{We assume a distance of 17~Mpc to the Virgo cluster}.
The Virgo cluster is dynamically young and spiral-rich.
The cluster spiral galaxies have lost up to $90$\,\% of their neutral hydrogen, i.e. they are
H{\sc i} deficient (Chamaraux et al. 1980, Giovanelli \& Haynes 1983).
Imaging H{\sc i} observations have shown that these galaxies have truncated H{\sc i} disks 
(Giovanelli \& Haynes 1983, Cayatte et al. 1990). Thus, the cluster environment changes the H{\sc i} content and morphology of 
Virgo cluster spiral galaxies. However, Virgo spiral galaxies are not or only mildly CO-deficient 
(Kenney \& Young 1989; see, however, Fumagalli et al. 2009).
In a recent H{\sc i} imaging survey of Virgo galaxies (VIVA: VLA Imaging of Virgo galaxies in Atomic gas, Chung et al. 2009), 
Chung et al. (2007) found seven spiral galaxies with long one-sided H{\sc i} tails. 
These galaxies are found in intermediate- to low-density regions (0.6-1 Mpc in projection from M87). 
The tails are all pointing roughly away from M87, suggesting that these tails are due to ram pressure stripping. 
Therefore, ram pressure stripping already begins to affect spiral galaxies around the cluster Virial radius. 
The further evolution of a galaxy depends critically on its orbit (see, e.g., Vollmer et al. 2001), 
i.e. a highly eccentric orbit will lead the galaxy at a high velocity into the cluster core, where the intracluster medium is 
densest and ram pressure will be very strong.

NGC~4330 is one of the Virgo spiral galaxies with a long H{\sc i} tail observed by Chung et al. (2007).
It has a maximum rotation velocity of $v_{\rm rot} \sim 180$~km\,s$^{-1}$ and a total H{\sc i} mass of
$M_{\rm HI}=4.5 \times 10^{8}$~M$_{\odot}$ (Chung et al. 2009). It is located at a projected distance of
$\sim 2^{\circ}$ (600~kpc) from the cluster center, i.e. it is relatively close to M~87, and has
a radial velocity of 400~km\,s$^{-1}$ with respect to the Virgo mean.   
The H{\sc i} deficiency of NGC~4330 is $0.8$ (Chung et al. 2007).
The H{\sc i} distribution in the galactic disk is truncated at about half the optical radius. In addition, NGC~4330 is one
of the rare Virgo galaxies showing an extended UV tail (Abramson et al. 2011). 
The H{\sc i} and UV tails show a significant offset, with the H{\sc i} tail being downwind of the UV tail.
At the leading edge of the interaction, the H$\alpha$ emission and dust extinction distribution is bent sharply
out of the galactic disk. These features are signs of active ram pressure stripping. 
For a comprehensive description of NGC~4330's optical, UV, and H{\sc i} data see Abramson et al. (2011).
These authors suggest a scenario in which NGC 4330 is falling into the cluster center for the first time and has experienced 
a significant increase in ram pressure over the last 200-400~Myr.

In the ram pressure stripping time sequence of Virgo cluster galaxies, Vollmer (2009) has classified NGC~4330
as a case of pre-peak ram pressure stripping together with NGC~4501.
This evolutionary stage is consistent with a galaxy orbit leading the galaxy to its closest
approach (620~kpc) of the cluster center (M~87) in $\sim 100$~Myr.

In this article we present new CO and 6~cm radio continuum observations of NGC~4330 and
give the details of the numerical simulations and give an explanation for the observed offset
between the H{\sc i} and UV tails, which is due to the decoupling of dense collapsing gas clouds from the
ram pressure wind. 
The article is structured in the following way:
the dynamical model is described in Sec.~\ref{sec:model} with the best fit model presented in Sec.~\ref{sec:bestfit}.
The CO(2-1) and 6~cm radio continuum observations are presented in Sec.~\ref{sec:molecular} and \ref{sec:radiocontinuum}.
The distribution of the polarized radio continuum emission is compared to MHD simulations in Sec.~\ref{sec:mhd}.
The peculiar H$\alpha$ and UV emission distributions of NGC~4330 are compared to the model in Sec.~\ref{sec:starformation}
and \ref{sec:sfr100}, followed by the overall picture of the ram pressure stripping event (Sec.~\ref{sec:overall}).
The stellar age distributions of the gas-free parts of galactic disk are presented in Sec.~\ref{sec:sfrhist}.
Finally, we give our conclusions in Sec.~\ref{sec:conclusions}.

\section{Dynamical model \label{sec:model}}

We use the N-body code described in Vollmer et al. (2001) which consists of 
two components: a non-collisional component
that simulates the stellar bulge/disk and the dark halo, and a
collisional component that simulates the ISM.
A new scheme for star formation has been implemented, where stars are formed
during cloud collisions and are then evolved as non-collisional particles.

\subsection{Halo, stars, and gas} 

The non--collisional component consists of 81\,920 particles, which simulate
the galactic halo, bulge, and disk.
The characteristics of the different galactic components are shown in
Table~\ref{tab:param}.
\begin{table}
      \caption{Total mass, number of particles $N$, particle mass $M$, and smoothing
        length $l$ for the different galactic components.}
         \label{tab:param}
      \[
         \begin{array}{lllll}
           \hline
           \noalign{\smallskip}
           {\rm component} & M_{\rm tot}\ ({\rm M}$$_{\odot}$$)& N & M\ ({\rm M}$$_{\odot}$$) & l\ ({\rm pc}) \\
           \hline
           {\rm halo} & 1.7 \times 10$$^{11}$$ & 32768 & $$5.0 \times 10^{6}$$ & 1200 \\
           {\rm bulge} & 5.7 \times 10$$^{9}$$ & 16384 & $$3.5 \times 10^{5}$$ & 180 \\
           {\rm disk} & 2.8 \times 10$$^{10}$$ & 32768 & $$8.7 \times 10^{5}$$ & 240 \\
           \noalign{\smallskip}
        \hline
        \end{array}
      \]
\end{table}
The resulting rotation velocity is $\sim$180~km\,s$^{-1}$ and the rotation curve
becomes flat at a radius of about 5~kpc. 

We have adopted a model where the ISM is simulated as a collisional component,
i.e. as discrete particles which possess a mass and a radius and which
can have inelastic collisions (sticky particles).
Since the ISM is a turbulent and fractal medium (see e.g. Elmegreen \& Falgarone 1996),
it is neither continuous nor discrete. The volume filling factor of the warm and cold phases
is smaller than one. The warm neutral and ionized gas fill about $30-50\%$ of the volume,
whereas cold neutral gas has a volume filling factor smaller than 10\% (Boulares \& Cox 1990). 
It is not clear how this fraction changes, when an external 
pressure is applied. In contrast to smoothed particles hydrodynamics (SPH), which is a 
quasi continuous approach and where the particles cannot penetrate each other, our approach 
allows a finite penetration length, which is given by the mass-radius relation of the particles.
Both methods have their advantages and their limits.
The advantage of our approach is that ram pressure can be included easily as an additional
acceleration on particles that are not protected by other particles (see Vollmer et al. 2001).

The 20\,000 particles of the collisional component represent gas cloud complexes which are 
evolving in the gravitational potential of the galaxy.
The total assumed gas mass is $M_{\rm gas}^{\rm tot}=4.3\,10^{9}$~M$_{\odot}$,
which corresponds to the total neutral gas mass before stripping, i.e.
to an \HI\ deficiency of 0, which is defined as the logarithm of the ratio between
the \HI\ content of a field galaxy of same morphological type and diameter
and the observed \HI\ mass.
To each particle a radius is attributed depending on its mass. 
During the disk evolution the particles can have inelastic collisions, 
the outcome of which (coalescence, mass exchange, or fragmentation) 
is simplified following Wiegel (1994). 
This results in an effective gas viscosity in the disk. 

As the galaxy moves through the ICM, its clouds are accelerated by
ram pressure. Within the galaxy's inertial system its clouds
are exposed to a wind coming from a direction opposite to that of the galaxy's 
motion through the ICM. 
The temporal ram pressure profile has the form of a Lorentzian,
which is realistic for galaxies on highly eccentric orbits within the
Virgo cluster (Vollmer et al. 2001).
The effect of ram pressure on the clouds is simulated by an additional
force on the clouds in the wind direction. Only clouds which
are not protected by other clouds against the wind are affected.
Since the gas cannot develop instabilities, the influence of turbulence 
on the stripped gas is not included in the model. The mixing of the
intracluster medium into the ISM is very crudely approximated by a finite
penetration length of the intracluster medium into the ISM, i.e. until this penetration
length the clouds undergo an additional acceleration due to ram pressure.

The particle trajectories are integrated using an adaptive timestep for
each particle. This method is described in Springel et al. (2001).
The following criterion for an individual timestep is applied:
\begin{equation}
\Delta t_{\rm i} = \frac{20~{\rm km\,s}^{-1}}{a_{\rm i}}\ ,
\end{equation}
where $a_{i}$ is the acceleration of the particle i.
The minimum value of $t_{\rm i}$ defines the global timestep used 
for the Burlisch--Stoer integrator that integrates the collisional
component. It is typically a few $10^{4}$~yr.

\subsection{Star formation \label{sec:sfr}}

We assume that the star formation rate is proportional to the cloud collision rate.
During the simulations stars are formed in cloud-cloud collisions. At each collision
a collisionless particle is created which is added to the ensemble of collisional and
collisionless particles. The newly created collisionless particles have zero mass
(they are test particles) and the positions and velocities of the colliding clouds after the collision. 
These particles are then evolved passively with the whole system. 
Since in our sticky-particle scheme there is mass exchange, coalescence, or fragmentation at the
end of a collision, the same clouds will not collide infinitely.
The local collision rate traces the cloud density and the velocity dispersion of the collisional component.
Since the cloud density increases rapidly with decreasing galactic radius, the number of newly created particles 
rises steeply toward the galaxy center.
To limit the total number of
these particles without losing information in the region of interest, we limit
the creation of collisionless star particles to galactocentric radii larger than 3.5~kpc.
This is substantially smaller than the gas stripping radius at the timestep of interest
which is $\sim 5$~kpc.

The information about the time of creation is attached to each newly created star particle.
In this way the H$\alpha$ emission distribution can be modeled by the distribution of
star particles with ages less than $10$~Myr. The UV emission of a star particle in the two GALEX bands
is modeled by the UV flux from single stellar population models from STARBURST99 (Leitherer et al. 1999).
The total UV distribution is then the extinction-free distribution of the UV emission of the newly created
star particles. For the unperturbed galaxy the resulting power law between star formation based on the model UV emission 
and the total/molecular gas surface density (see Sect.~\ref{sec:comodel}) has an exponent of $1.7$/$1.2$, respectively
(Fig.~\ref{fig:schmidtlaw}). This is close to the observational findings of Bigiel et al. (2008).
The simulations start $600$~Myr before peak ram pressure.

\section{Search for the best fit model \label{sec:bestfit}}

In this section we will constrain the parameters of the ram pressure stripping event
which are (i) the peak ram pressure, (ii) the temporal ram pressure profile,
(iii) time since peak ram pressure, (iv) the inclination angle between the galaxy's 
disk and the intracluster medium wind direction, and (v) the azimuthal viewing angle for 
the observed inclination and position angles.
These parameters are related to the observed quantities which are (1) the position angle,
(2) the inclination angle of the galactic disk, (3) the line-of-sight velocity
of the galaxy with respect to the cluster mean, and (4) the projected ICM wind direction.
The position angle and inclination of NGC~4330 define a plane in three dimensional space.
The model galaxy can then be rotated within this plane by the azimuthal viewing angle (see below). 
The three dimensional model wind direction, the line-of-sight velocity of the galaxy, and
the projected ICM wind direction are thus functions of the azimuthal viewing angle.

In the case of a smooth static ICM ram pressure is proportional to the ICM density
$\rho_{\rm ICM}$ and the square of the galaxy velocity with respect to the Virgo cluster 
$\vec{v_{\rm gal}}$. If the ICM is moving with respect to the cluster mean velocity
the expression for ram pressure yields:
\begin{equation}
p_{\rm ram}=\rho_{\rm ICM} (\vec{v_{\rm gal}}-\vec{v_{\rm ICM}})^{2}\ ,
\label{eq:icmwind}
\end{equation}
where $\vec{v_{\rm ICM}}$ is the velocity of the intracluster medium, 
and $\vec{v_{\rm gal}}$ is the galaxy velocity with respect to the Virgo cluster.
We assume a classical static intracluster medium, i.e. $v_{\rm ICM}=0$~km\,s$^{-1}$ and temporal 
ram pressure profile of the following form:
\begin{equation}
p_{\rm ram}=p_{\rm max} \frac{t_{\rm HW}^{2}}{t^{2}+t_{\rm HW}^{2}}\ ,
\label{eq:rps}
\end{equation}
where $t_{\rm HW}$ is the width of the profile (Vollmer et al. 2001).

We only consider model snapshots leading to a gas truncation radius of about half the optical radius
as it is observed.
We set $p_{\rm max}$=5000~cm$^{-3}$(km\,s$^{-1}$)$^{2}$ and $t_{\rm HW}$=100~Myr.
We define $t=0$~Myr as the time when ram pressure is maximum.

The ram pressure efficiency also depends on the inclination angle $i$
between the galactic disk and the ICM wind direction (Vollmer et al. 2001).
Since the H{\sc i} observations indicate that
stripping occurs more face-on, we do not consider edge-on stripping.
We made 2 simulations with 2 different 
inclination angles between the galaxy's disk and the ICM wind direction:
(i) $i=60^{\circ}$ and (ii) $i=75^{\circ}$.
An inclination of $i=0^{\circ}$ means that the galactic disk is
parallel to the ICM wind direction.

The last open parameter, the azimuthal viewing angle, is chosen
in a way to fit the observed H{\sc i} distribution and to reproduce
the positive line-of-sight component of the wind direction (the galaxy's radial velocity with
respect to the cluster mean is small). 
\begin{figure*}
	\resizebox{\hsize}{!}{\includegraphics{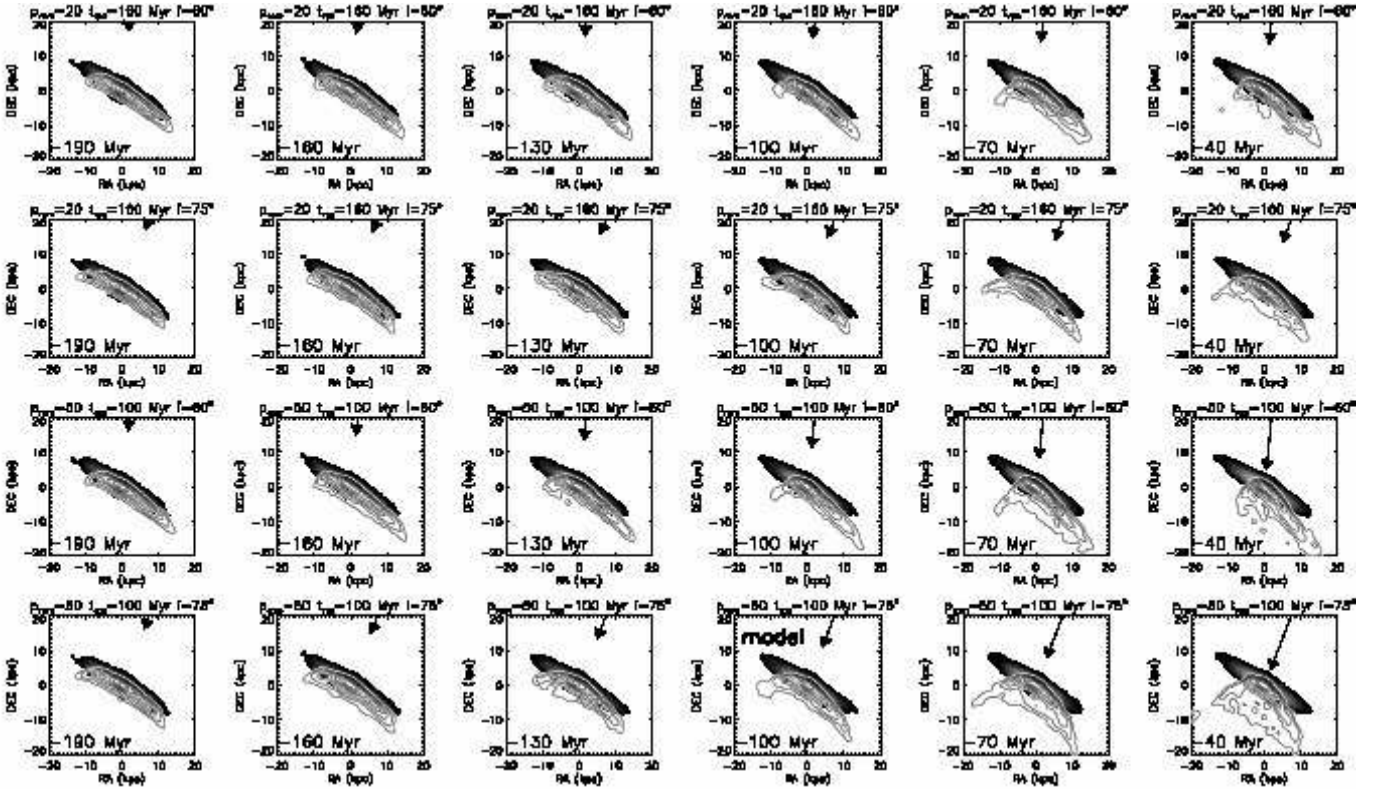}}
	\caption{Snapshots of the 4 ram pressure stripping simulations. The ram pressure
          maximum in units of 100~cm$^{-3}$(km\,s$^{-1}$)$^{2}$, the duration $t_{\rm HW}$,
          and the inclination angle between the orbital and galactic disk plane $i$
          are shown on top of each panel. The timestep with respect to the ram pressure
          maximum are shown in the lower left corner of each panel. The arrow indicate
          the direction of the ram pressure wind with its size proportional to ram pressure strength.
          The stellar disk is shown in grayscale with lower column densities being darker.
          The gas is shown as contours. The best-fit model ($p=5000$~cm$^{-3}$(km\,s$^{-1}$)$^{2}$,
          $t=-100$~Myr, $i=75^{\circ}$) is labeled with ``model''.
	} \label{fig:n4330_pathfinder_1}
\end{figure*} 

Snapshots of the 4 simulations are shown in Fig.~\ref{fig:n4330_pathfinder_1}.
Inclination angles between the orbital and galactic planes smaller than $75^{\circ}$ do not produce
a gas tail which significantly bends out of the galactic plane on the lower left side of the galaxy.
The gas tail for $i=75^{\circ}$ reaches the observed morphology between $t=-100$ and $t=-70$~Myr, i.e. 
100/70~Myr before peak ram pressure or closest approach to the cluster center. 
Based on the H{\sc i} morphology we chose $p=5000$~cm$^{-3}$(km\,s$^{-1}$)$^{2}$ and $t=-100$~Myr as the best fit model, 
because these parameters are most consistent with NGC~4330's orbit within the Virgo cluster (Vollmer 2009).
 
\begin{figure}
	\resizebox{\hsize}{!}{\includegraphics{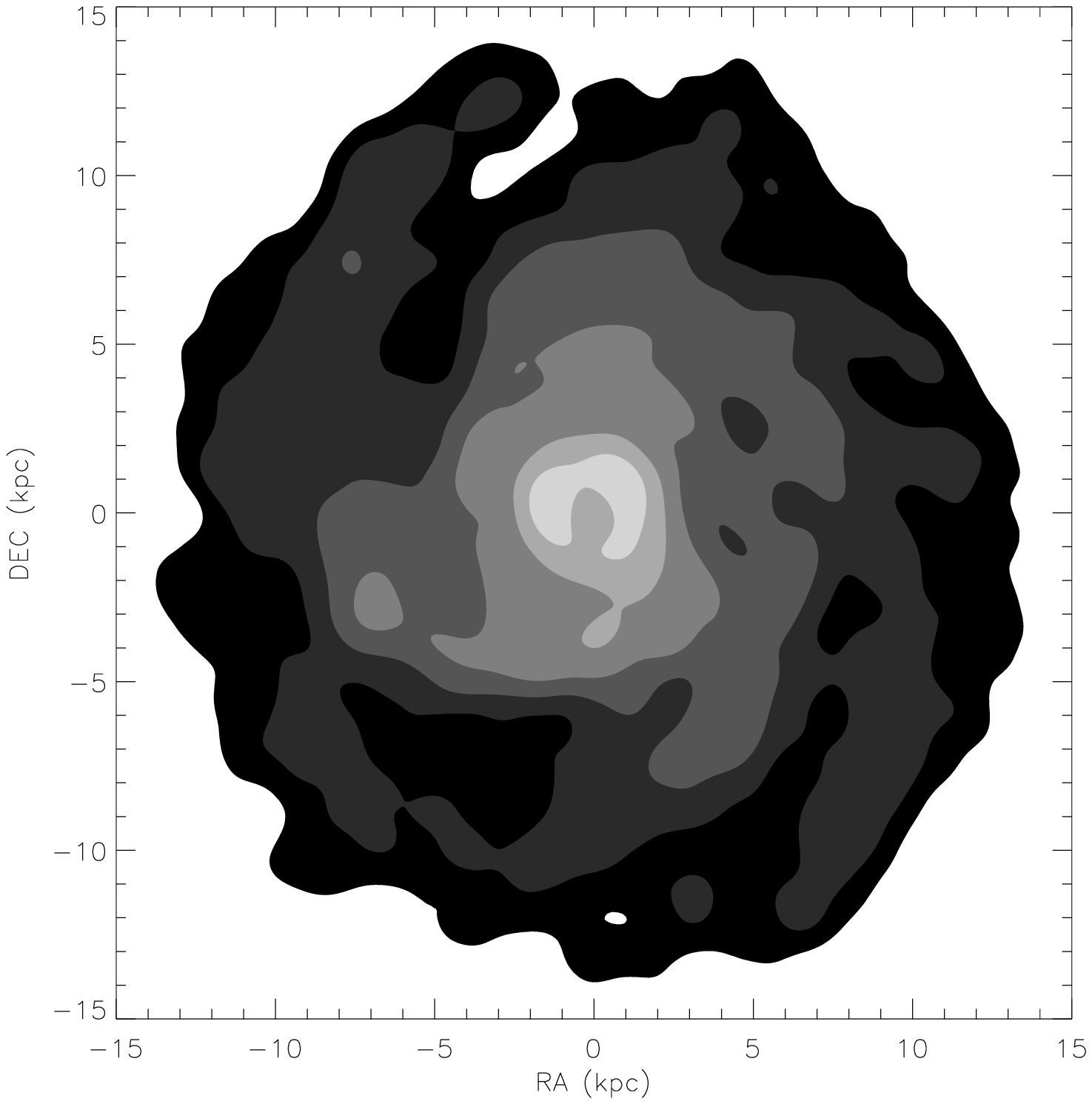}}
        \resizebox{\hsize}{!}{\includegraphics{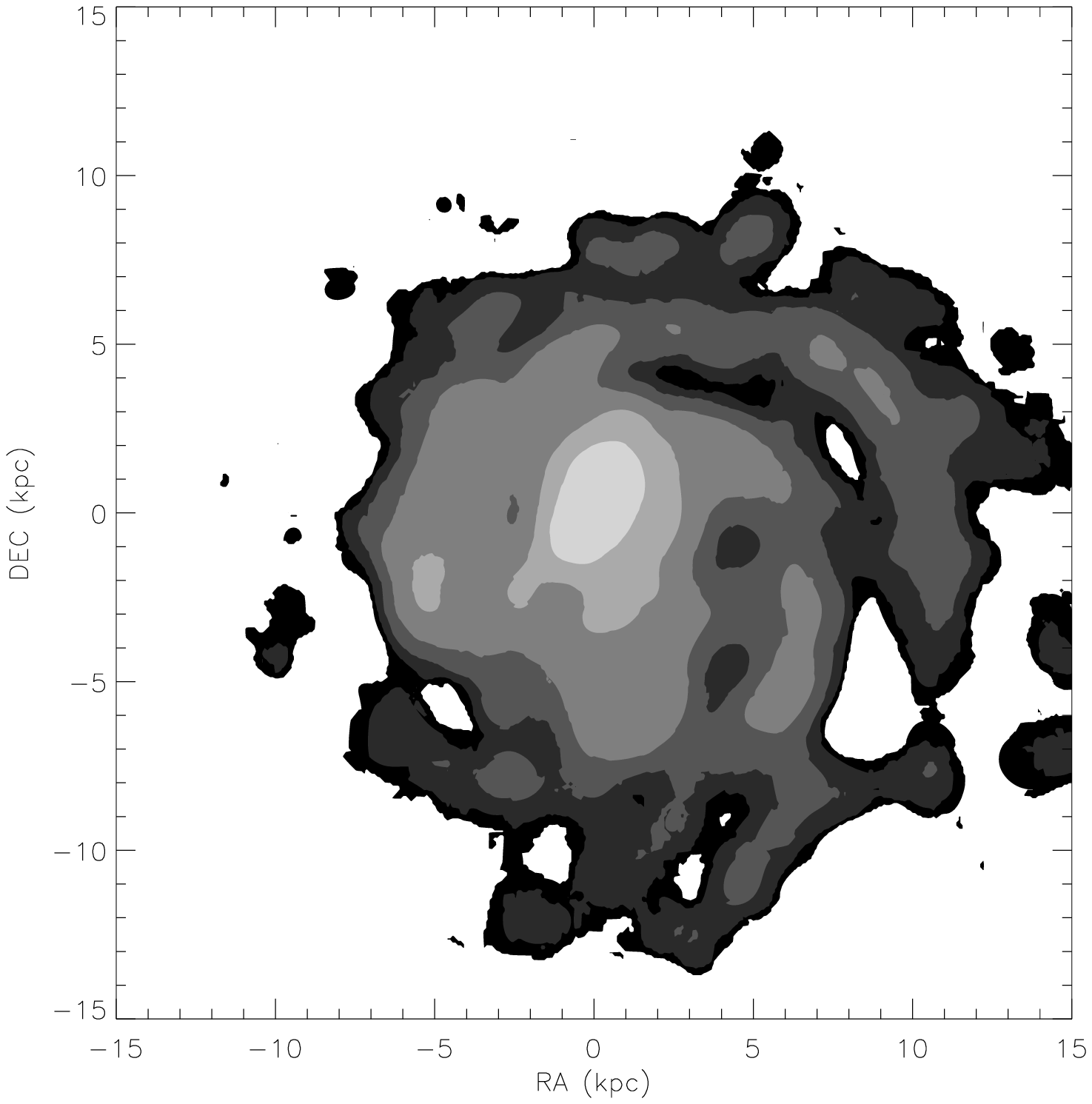}}
	\caption{Face-on model H{\sc i} distribution. Upper panel: initial condition ($t=-600$~Myr). Lower panel: 
	  best-fit model ($t=-100$~Myr). The galaxy moves to the left, i.e. the ram pressure wind
	  blows from the left. The low surface density gas at $RA < -8$~kpc is located at high latitudes,
	  i.e. it is extraplanar.
	  The contour levels are $(1,2,4,8,16,32,64) \times 10^{-19}$~cm$^{-2}$.
	} \label{fig:face-on}
\end{figure} 
The model galaxy develops a spiral structure (upper panel of Fig.~\ref{fig:face-on}) which influences gas stripping by ram pressure 
(Schulz \& Struck 2001).
In a first set of simulations we used different initial conditions where the spiral arms
are in different locations than in the present model.
By chance, a prominent spiral arm was located at the windward side of the galaxy at the timestep of interest.
Because of its higher column density a spiral arm is more resistant to ram pressure stripping.
The northeastern extraplanar gas thus had a much higher surface density.
Since this is not observed, we discarded these first simulations and chose different initial conditions
to avoid a spiral arm at the windward side at the timestep of interest ($-100$~Myr). 
In our best-fit model the major spiral arms of the outer disk are located on the trailing side (lower panel of Fig.~\ref{fig:face-on})).

\section{Comparison with H{\sc i} data \label{sec:comphi}}

\begin{figure*}
	\resizebox{15cm}{!}{\includegraphics[angle=-90]{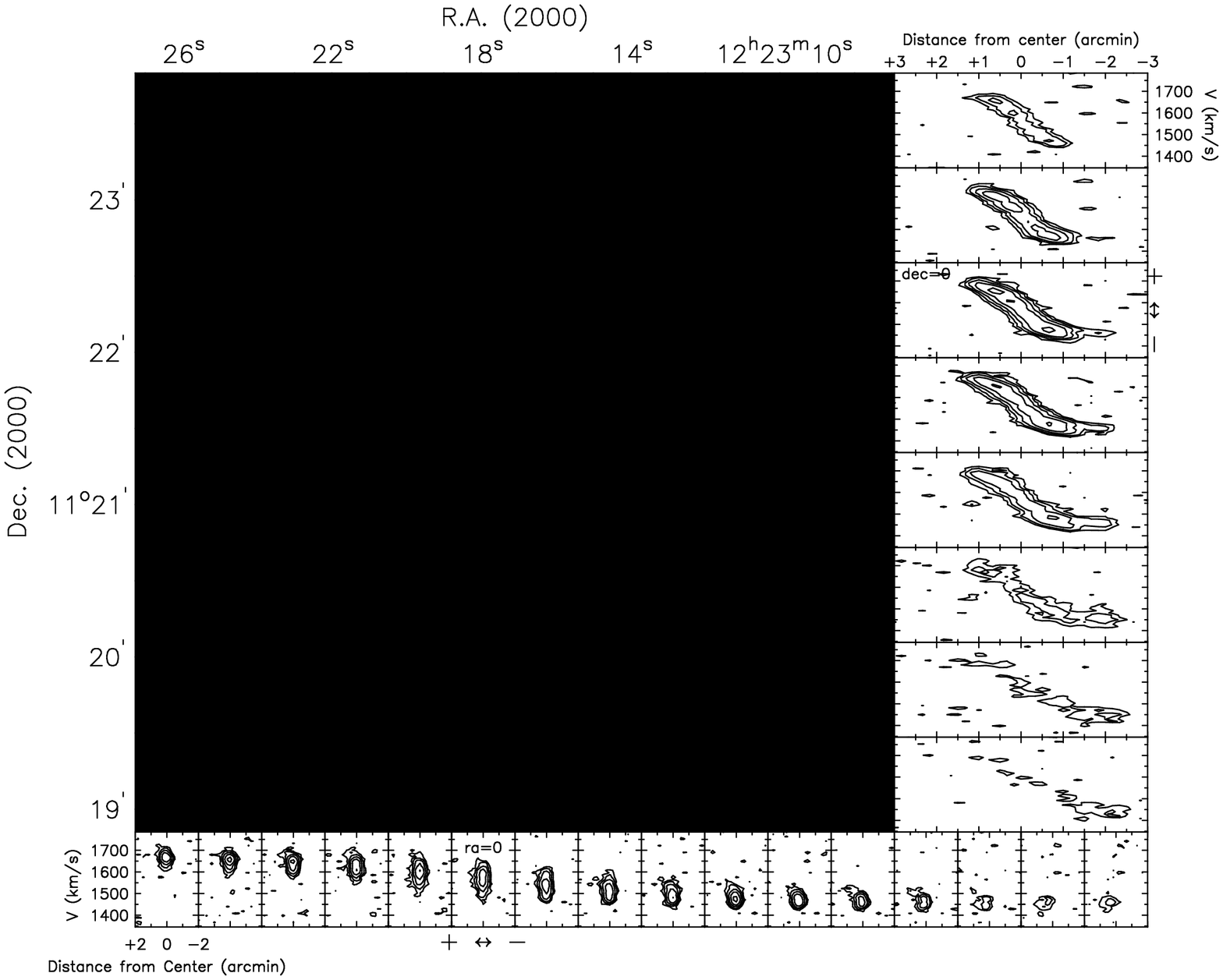}}
        \resizebox{15cm}{!}{\includegraphics[angle=-90]{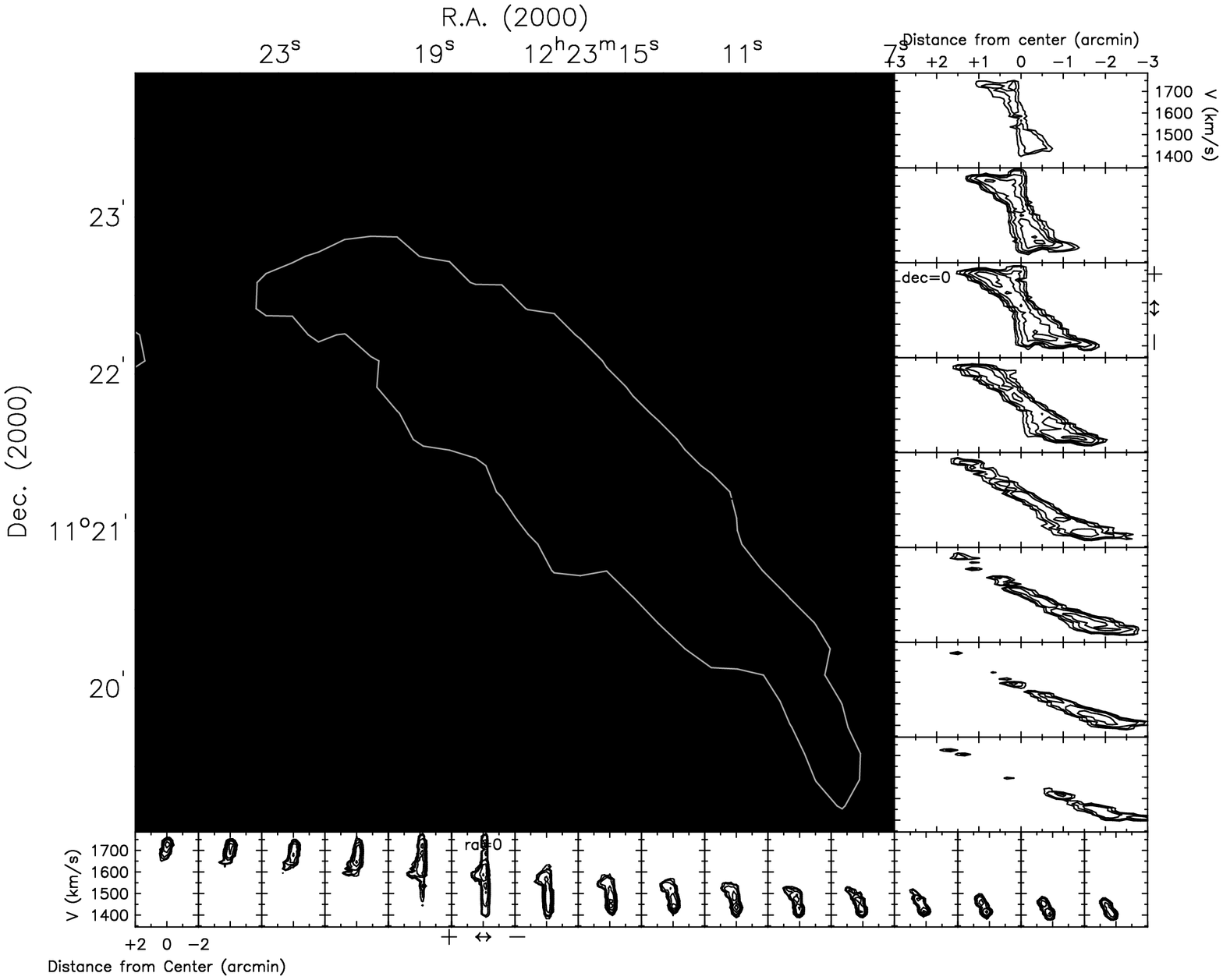}}
	\caption{Gas distributions and associated position velocity diagrams. Upper panel:
          H{\sc i} observations (grayscale and contours; Chung et al. 2009). 
	  The contour levels of the H{\sc i} distribution are
	  $(1,2,4,8,16,32,64) \times 10^{-19}$~cm$^{-2}$. The beam is shown in the lower left corner.
	  Lower panel: grayscale: observed H{\sc i} distribution. Contours: model atomic gas distribution.
          The relative contour levels are same as those of the H{\sc i} distribution. 
	} \label{fig:ngc4330.pvd}
\end{figure*}

A model cube (right ascension, declination, radial velocity) with the properties of
the H{\sc i} observations of Chung et al. (2009) is produced from the model gas particle distribution. 
The observed and model gas distributions and associated position velocity diagrams are shown in Fig.~\ref{fig:ngc4330.pvd}.
The model gas distribution reproduces qualitatively the main features of the observations:
(i) the truncation of the galactic gas disk to the northeast and southwest,
(ii) the gas distribution along the minor axis is larger on the shadowed side
(southeast), and (iii) the existence of the southwestern tail, especially there is a model counterpart (a kink)
to the discontinuity of the gas distribution between the disk and the tail gas.
Despite the overall agreement with observations, there are disagreements:
(i) the upturn to the southeast is more pronounced in the model compared to observations.
In particular, the low column density gas on the eastern side of the galaxy is more extended
than the observed gas distribution; (ii) the structure of the model tail has a pronounced
part which is bent by an angle of $18^{\circ}$ with respect to the galactic disk, whereas the observed 
tail bends by an angle of $\sim 30^{\circ}$; (iii) the observed vertical extent of the
H{\sc i} disk is about twice as large as that of the model gas disk.

The observed gas kinematics along the major axis are well reproduced by the model, except that the
model has an intrinsically steeper rotation curve and a somewhat high rotation velocity.
The extensions to high velocities observed in the position velocity diagrams of the gas tail are also present in the model.
However, there are differences between the observed and the model position velocity diagrams
along the minor axis:
(i) the observed position velocity diagram along the minor axis at the galaxy center and
to the northeast show the highest linewidths of $\sim 150$~km\,s$^{-1}$;
(ii) the corresponding model position velocity diagrams are two times broader;
(iii) in the upturn region (northeast) the observed velocities on the windward side
are, if they are different, higher than those of the opposite side, whereas the model velocities are lower.
Thus, whereas the model kinematics of the tail region are in qualitative agreement with observations, 
the model kinematics of the upturn region are different.

\section{The role of gas shadowing}

The main ingredients of a ram pressure stripping event are (i) ram pressure, (ii) galactic rotation, (iii)
gas shadowing, i.e. the finite penetration length of the intracluster medium into the ISM, and (iv) projection.

Our results can be compared to similar ram pressure configurations simulated with an SPH (Schulz \& Struck 2001)
and 3D hydrodynamical code (Roediger \& Br\"{u}ggen 2006). Projections similar to that of NGC~4330 can be found in
both articles. Fig.~10 of Schulz \& Struck (2001) shows a
galaxy stripped at an angle of $40^{\circ}$ with respect to the galactic plane. The stripped gas at the windward side has
a relatively high column density and is distributed approximately orthogonally to the wind direction.
The gas tail of the trailing side has also a relatively high column density and is shifted by an angle of 
$\sim 10^{\circ}$ with respect to the galactic plane.
The gas distribution is similar to that of our simulations with ($i=60^{\circ}$, $p_{\rm ram}=20$, $t=-130$ to $-100$~Myr) or 
($i=60^{\circ}$, $p_{\rm ram}=50$, $t=-160$ to $-130$~Myr).

The upper left panel of Fig.~5 of Roediger \& Br\"{u}ggen (2006), where the galaxy is stripped at an angle of
$60^{\circ}$ with respect to the galactic plane, can also be compared to our simulations.
The extraplanar gas at the trailing side is aligned with the wind direction, that of the leading side
is at an angle of $45^{\circ}$. Whereas the behavior of the leading side is similar to that of
our simulations with ($i=75^{\circ}$, $p_{\rm ram}=50$, $t=-130$ to $-70$~Myr), the trailing side of our 
(and the SPH) model is always much closer to the galactic disk than in the 3D hydrodynamical simulations.
The more efficient stripping of the trailing side is due to Kelvin-Helmholtz instabilities creating turbulent motions and mixing in
the extraplanar leeward side which makes ISM stripping more efficient. 

We thus conclude that the angle between the
stripped gas at the leeward side of the galaxy with respect to the galactic disk depends on
behavior of the stripped gas involving instabilities and mixing. If the viscosity of the intracluster medium
is high, the stripped gas shows less structure and turbulence (Roediger \& Br\"{u}ggen 2008).
Observations as those of NGC~4330 or NGC~4522 (Vollmer et al. 2008) have to constrain the role of instabilities on the stripped gas on the
leeward side of the galaxy. In terms of our simple model, the influence of instabilities and mixing is
less shadowing or an increase of the penetration length of the intracluster medium.
In addition, the gas surface density might decrease. The latter effect is not included in our model. 
The case of NGC~4522 (Vollmer et al. 2008) shows that gas shadowing is needed.

\begin{figure*}
        \resizebox{15cm}{!}{\includegraphics[angle=-90]{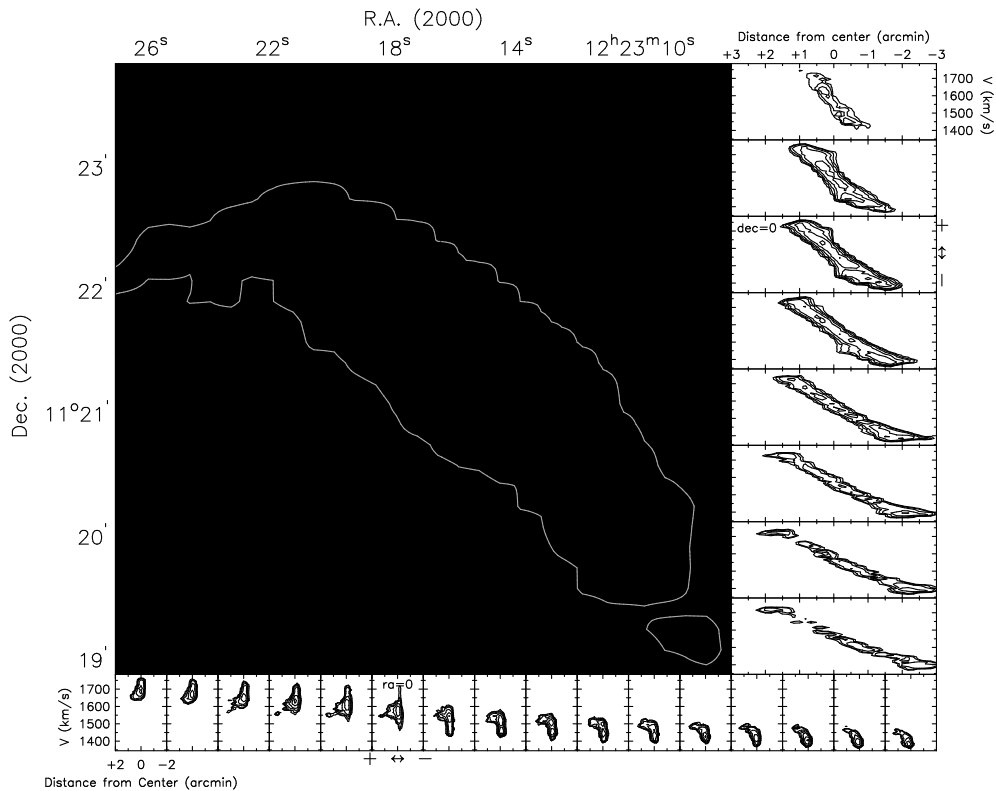}}
	\caption{Model without gas shadowing.
	  The relative contour levels are same as those of the H{\sc i} distribution in Fig.~\ref{fig:ngc4330.pvd}.
	} \label{fig:ngc4330.pvd.old}
\end{figure*}

To investigate the effect of gas shadowing, we have redone our ``best-fit'' simulation without gas shadowing.
In order to obtain the same gas morphology $100$~Myr before peak ram pressure, we had to adjust the temporal 
ram pressure profile: $p_{\rm max}=2000$~cm$^{-3}$(km\,s$^{-1})^{2}$ and $t_{\rm HW}=160$~Myr.
Thus, in the simulation without gas shadowing a two times smaller ram pressure is needed to obtain the same
effect as in the simulation including shadowing.
The resulting atomic gas distribution at $t=-100$~Myr is shown in Fig.~\ref{fig:ngc4330.pvd.old}.
The gas distribution is more extended to the south than that of the simulation including gas shadowing. 
The main difference between the two simulations is the shape of the southwestern gas tail:
in the simulation with shadowing there is (i) a kink between the
disk and the tail and (ii) the high surface density part of the tail which is bent by a
different angle with respect to the galactic disk.
Without shadowing the gas tail is bent by a constant angle of $\sim 30^{\circ}$, with shadowing
this angle is $18^{\circ}$.

The gas kinematics along the minor axis of the two simulations are similar.
Since without shadowing all gas clouds are pushed by ram pressure, 
the velocities along the major axis reach lower values in the simulations without than with gas shadowing.
The position velocity diagrams along the major axis of the simulations without shadowing reproduce
better the observed velocity diagrams than those of the simulations with shadowing.

Based on the comparison between the simulations with and without gas shadowing we suggest that
(i) the discontinuity of the observed H{\sc i} distribution between the disk and the tail could be due
to shadowing and (ii) the effect of shadowing might decrease at the end of the gas tail.
In other words, the penetration length of the intracluster medium into the tail ISM increases with 
increasing distance from the galaxy center. This is expected if turbulent mixing of the
intracluster medium into the ISM becomes important (Roediger \& Br\"{u}ggen 2006).

In the following we only consider the simulations with gas shadowing. The further results and conclusions
of our work do not depend on gas shadowing.

\section{Molecular gas content \label{sec:molecular}}

\subsection{Observations}

The observations of the CO(2--1) line, with rest frequency of 230.53799 GHz, 
were carried out at the 30~meter  millimeter-wave
telescope on Pico Veleta (Spain) run by the Institut de RadioAstronomie
Millim\'etrique (IRAM).  The CO(2--1) observations used the HERA  multi-beam
array, with 3 $\times$ 3 dual-polarization receivers, and the WILMA  autocorrelator backend
with 2MHz spectral resolution.  The spatial and spectral resolutions are $11''$ and $2.6$~km\,s$^{-1}$.  
The HERA observations were made in February and March  2006 and the
CO(1--0) in December 2008.
In both cases, a nutating secondary ("wobbler") was used with a throw  of 180--200 arcseconds
in order to be clear of any emission from the galaxy.

\begin{figure}
	\resizebox{\hsize}{!}{\includegraphics{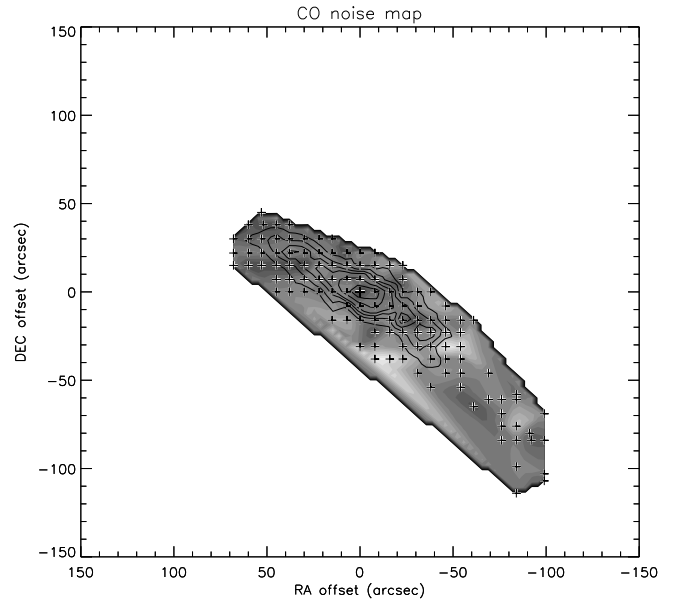}}
	\caption{Map of HERA CO(2--1) pointings and the relative noise in the summed
          spectra. The gray scale image of the relative noise has been interpolated using
	  tesselation. Darker regions correspond to lower noise levels. Contours:
          CO(2--1) emission distribution.
	} \label{fig:n4330_co_noisemap}
\end{figure}

Data reduction was straightforward, eliminating any obviously bad channels
and excluding the spectra taken under particularly poor conditions (system temperature
over 1000~K).  Spectra were then summed position by position.
System temperatures of the final spectra ranged from 200 to 500~K on the Ta$^*$ scale.
In Fig.~\ref{fig:n4330_co_noisemap} we present the pointings and the map of relative noise of
the summed spectra.
The telescope main-beam and forward efficiencies are $\eta_{\rm mb}=0.54$ and $\eta_{\rm for}=0.90$.
At the assumed distance of NGC~4330, 17~Mpc, the CO(2--1) beam corresponds to
0.9~kpc.  In order to convert CO integrated intensities into molecular gas masses,
we have assumed a $N({\rm H}_2) / I_{\rm CO(2-1)}$ ratio 
of $2 \times 10^{20}$~H$_2$ mol~cm$^{-2}$ per K kms$^{-1}$.
Our conclusions, however, do not depend strongly on the $N({\rm H}_2) / I_{\rm CO}$ ratio within
reasonable variations.
We made a first moment map by interpolating the CO(2--1) flux density of the pointings
on a regular grid as we did in Vollmer et al. (2006).

\subsection{Results \label{sec:comodel}}

\begin{figure}
	\resizebox{\hsize}{!}{\includegraphics{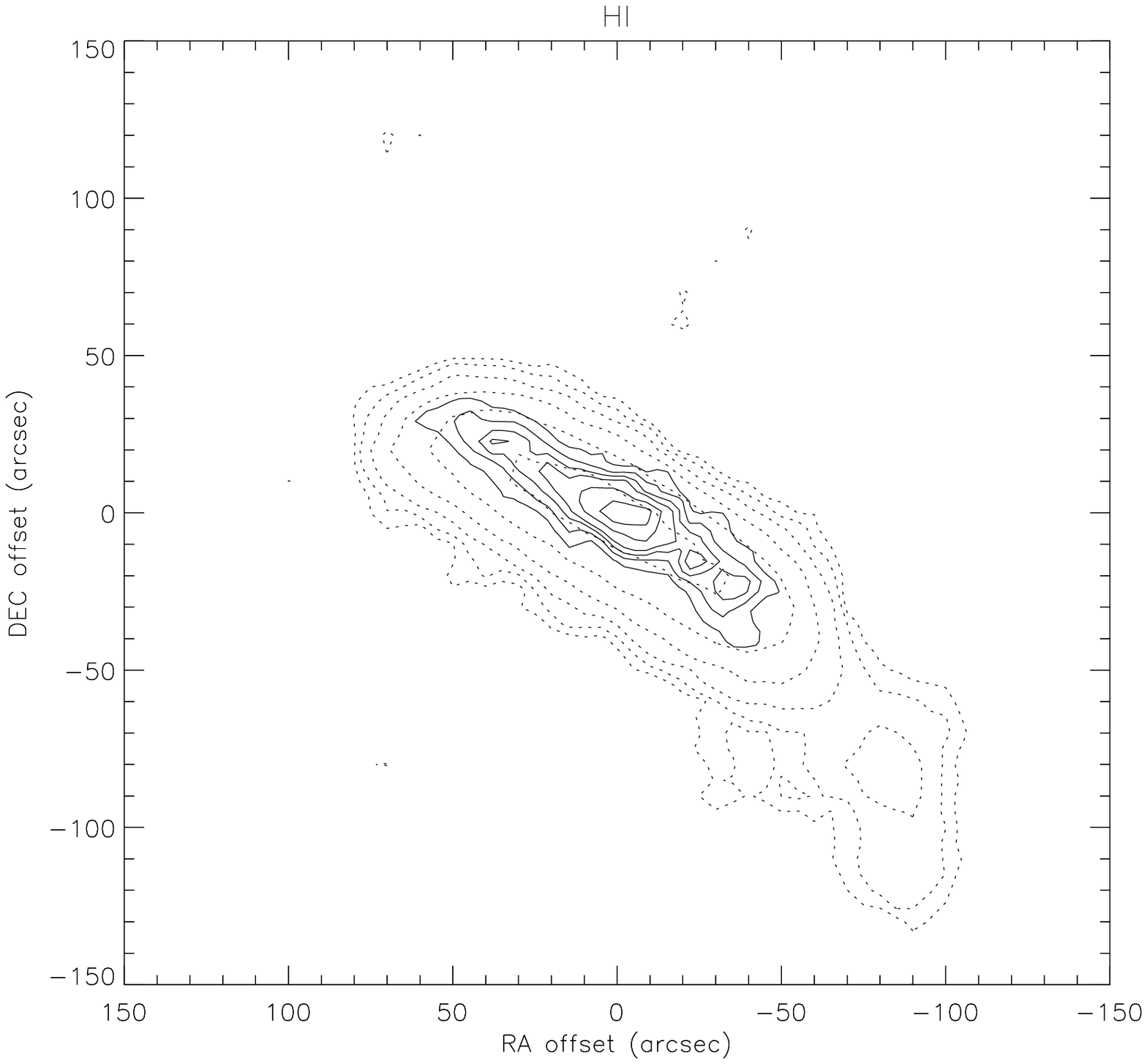}}
	\resizebox{\hsize}{!}{\includegraphics{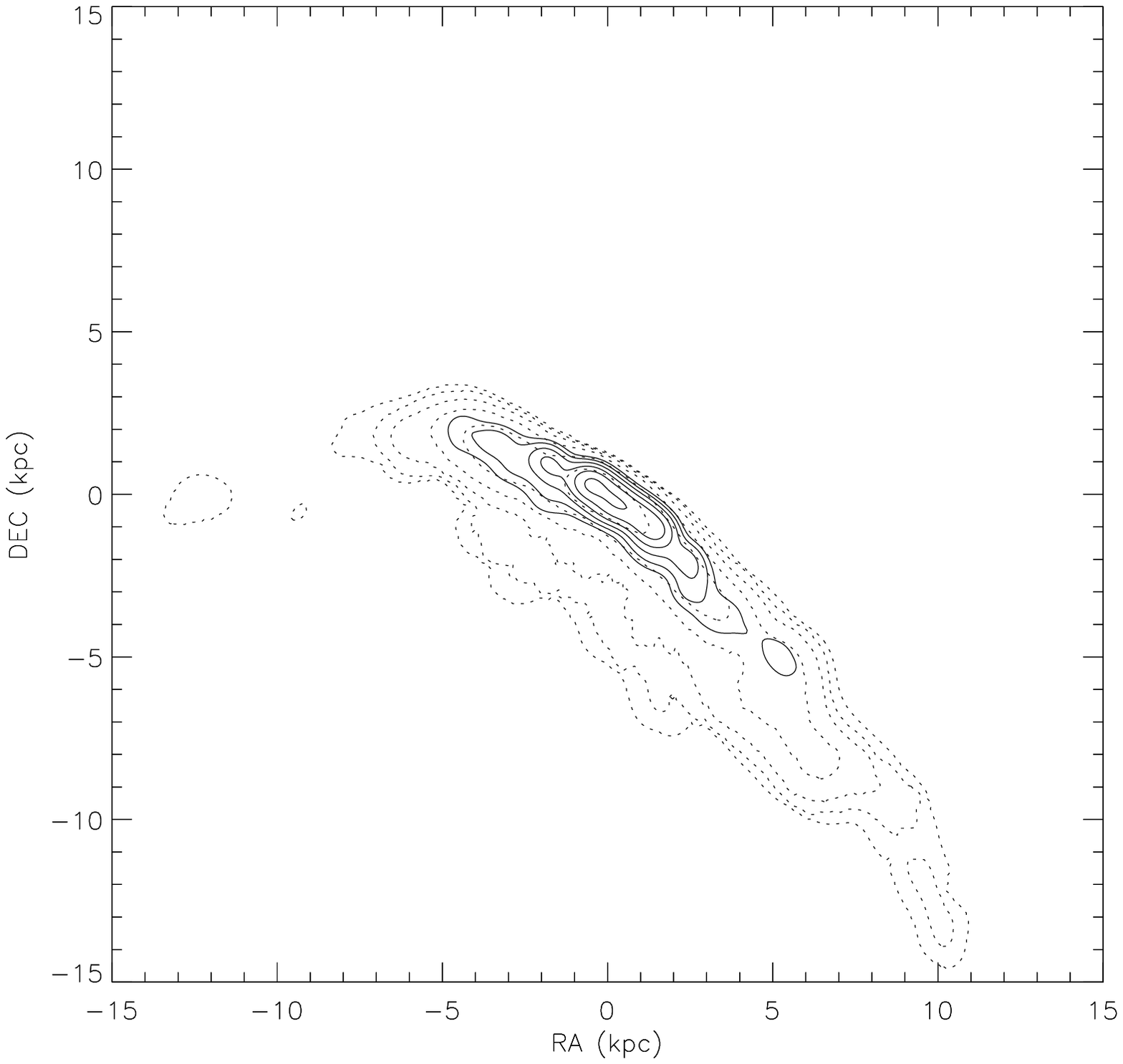}}
	\caption{Molecular CO(2--1) gas distribution of NGC~4330 (solid contours).
        Dotted contours: H{\sc i} column density distribution.
        Upper panel: observations. Lower panel: model.
	} \label{fig:n4330_hi_co}
\end{figure}

The distribution of molecular gas is shown in the upper panel of Fig.~\ref{fig:n4330_hi_co}.
The morphology consists of a strong maximum in the galactic center and an asymmetric molecular gas disk.
The slight bending of the molecular gas distribution in the galactic disk is reminiscent of spiral structure.
To the southwest, the molecular gas disk ends well within the H{\sc i} distribution, whereas it
extends to the edge of the H{\sc i} distribution in the northeast.
We observe a faint extension in the upturn region. We do not detect any extraplanar CO emission in the tail region
down to a gas mass of $\sim 6 \times 10^{6}$~M$_{\odot}$ (including helium).
We also checked for faint CO emission in the gas-free galactic disk on the leading side
of the interaction (northeast). No CO emission is found to a level of $\sim 2 \times 10^{6}$~M$_{\odot}$
when 15 spectra are averaged.
In contrast, we found $\sim 2 \times 10^{6}$~M$_{\odot}$ of molecular gas in the gas-free galactic disk on the leading side
of the interaction in NGC~4522 (Vollmer et al. 2006; see Sec.~\ref{sec:overall}).
If such a gas mass were present in the upturn region of NGC~4330, we should have detected it.
To obtain the model molecular gas distribution of NGC~4330 we follow Vollmer et al. (2006) and weight
the particle mass with $1.4 \sqrt{\rho}$, where $\rho$ is the large-scale gas density.
The model molecular gas distribution is shown in the lower panel of Fig.~\ref{fig:n4330_hi_co}.
The model molecular gas surface density distribution has a central maximum. The molecular gas disk
is more extended to the northeast than to the southwest. This morphology is in close agreement with
our observations. However, both sides of the extended model molecular gas disk are slightly
bent to the south, a behavior which is not clearly observed.
In agreement with observations, no molecular gas is found in the extraplanar southwestern gas tail.

\section{Radio continuum emission at 6~cm \label{sec:radiocontinuum}}

\subsection{Observation}

NGC~4330 was observed at 4.85~GHz during 6~h 40~min on December 1 and 3, 2009
with the Very Large Array (VLA) of the National
Radio Astronomy Observatory (NRAO)\footnote{NRAO is a facility of
National Science Foundation operated under cooperative agreement by
Associated Universities, Inc.} in the D array configuration. The
band passes were $2\times 50$~MHz. We used 3C286 as the flux
calibrator and 1254+116 as the phase calibrator, the latter of which
was observed every 40~min. Maps were made for both wavelengths using
the AIPS task IMAGR with ROBUST=3. The final cleaned maps were
convolved to a beam size of $18'' \times 18''$.
The rms levels of the polarized emission is 13~$\mu$Jy.
We obtain apparent B vectors by rotating the observed E vector by $90^{\circ}$,
uncorrected for Faraday rotation.

\subsection{Results}

\begin{figure}
	\resizebox{\hsize}{!}{\includegraphics{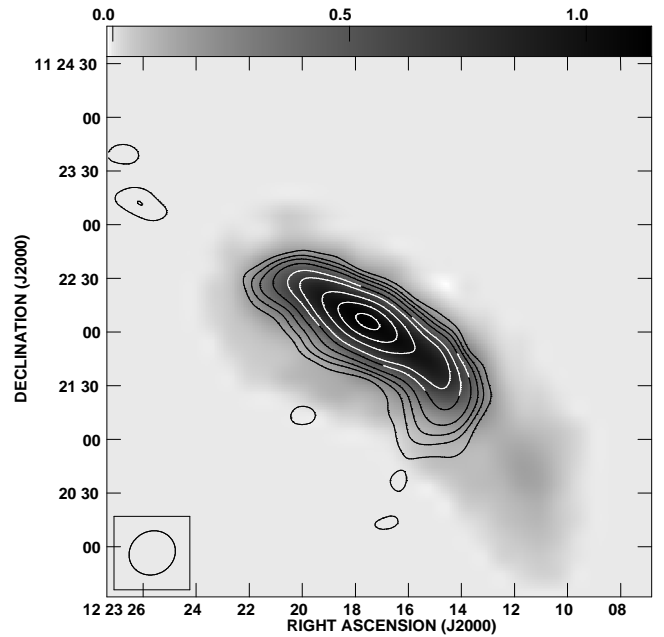}}
	\caption{Total power radio continuum emission at 6~m (contour) and the H{\sc i} distribution 
	  (grayscale in Jy/beam\,km\,s$^{-1}$). 
	  Contour levels are $(3,5,7,10,15,20,30,40,50) \times 20$~$\mu$Jy.
	} \label{fig:n4330_cont_hi}
\end{figure}

The observed radio continuum emission on the H{\sc i} column density distribution is shown in 
Fig.~\ref{fig:n4330_cont_hi}.
The upturn observed in H$\alpha$/UV (Fig.~14 of Abramson et al. 2011) and CO is clearly visible in
the total power emission. Moreover, the total power emission shows an asymmetry with respect to the major axis.
The emission is more extended on the southeastern downwind side than on the northwestern
windward side. The extraplanar H{\sc i} tail is not detected at 6~cm. However, the total emission
bends strongly to the south at the western border of the galactic gas disk. This behavior is
not visible at any other wavelength except at $20$~cm (Fig.~20 of Abramson et al. 2011).
We can only speculate that the cosmic ray gas, having a low column density, is stripped more
efficiently than the neutral gas. This effect has already been suggested by Vollmer et al.
(2009) for NGC~4438 and Crowl et al. (2005) in NGC~4402. 
However, one would expect a compression of the comic ray gas, which is 
in contradiction with the absence of polarization in this region. Alternatively, there might be 
enhanced diffusion of comic ray electrons into the tail region due to, e.g., an outflow.

\begin{figure}
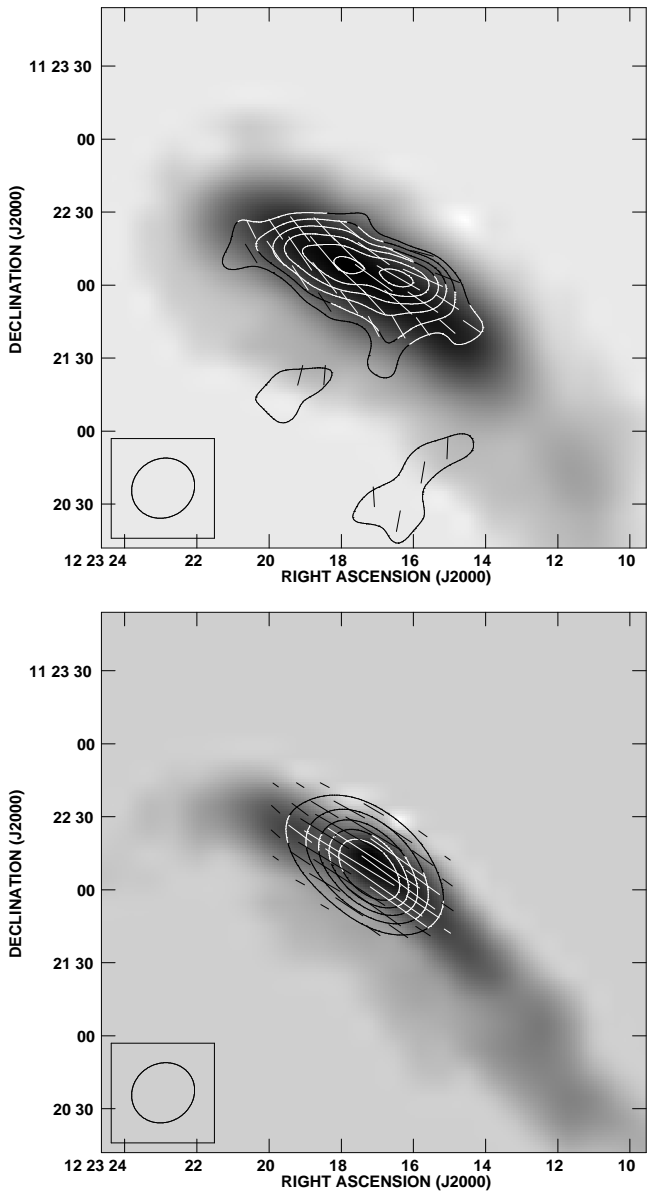

	\resizebox{\hsize}{!}{\includegraphics{n4330_hi_pi_obs.ps}}
	\resizebox{\hsize}{!}{\includegraphics{n4330_hi_pi_model.ps}}
	\caption{Polarized radio continuum emission (contours) and apparent magnetic field vectors
         (lines) on the H{\sc i} distribution (grayscale). Upper panel:
          observations; contour levels are $(3,4,5,6,7,8,10) \times 12$~$\mu$Jy.
          Lower panel: model; contour levels are $(3,4,5,6,7,8,10)$ times an arbitrary level.
	} \label{fig:n4330_hi_pi}
\end{figure}

The observed polarized radio continuum emission together with the magnetic field vectors
on the H{\sc i} column density distribution are shown in the upper panel of 
Fig.~\ref{fig:n4330_hi_pi}.
The polarized radio continuum emission is mainly found in the inner part of the gas disk.
Its total extent along the galaxy's major axis is $\sim 1'$.
The emission is almost symmetric with respect to the major and minor axis. 
Two maxima are observed which follow the spiral arms detected in CO
(Fig.~\ref{fig:n4330_hi_co}). We do not observe
an asymmetric ridge of polarized radio continuum emission as we did in all
other ram pressure affected Virgo spiral galaxies (Vollmer et al. 2007).
On the leading side of the interaction we observe a faint upturn of the polarized
radio continuum emission (12~23~20~+11~22~20) which is inside the H$\alpha$/total power radio continuum upturn.
The magnetic field vectors are not parallel to the galactic disk as one might 
expect it for an edge-on galaxy (see, e.g., Krause 2009). 
The magnetic field of the galaxy's western side has a characteristic X-structure, which is observed in 
edge-on spiral galaxies with a high star formation activity (e.g., NGC~253, Heesen et al. 2009).
The field in the northern and eastern part of the X is compressed by ram pressure. On the other hand, the magnetic field
in the east of the galaxy center is tilted clockwise by $\sim 30^{\circ}$ with respect to the disk plane.
There can be two reasons for this tilt: (i) bending of the galactic regular magnetic field or
(ii) enhanced Faraday rotation on the eastern side of the galactic disk. 
A rotation measure of $-130$~rad\,m$^{-2}$ is required to rotate the E vectors by $30^{\circ}$ which seems possible
(see below).

\subsection{MHD modeling \label{sec:mhd}}

Otmianowska-Mazur \& Vollmer (2003) studied the evolution of the
large-scale magnetic field during a ram pressure stripping event.
They calculated the magnetic field structure by solving the induction equation 
on the velocity fields produced by the dynamical model.
The polarized radio-continuum emission has been calculated by assuming
a Gaussian spatial distribution of relativistic electrons. This procedure
allowed them to study the evolution of the observable polarized radio
continuum emission during a ram pressure stripping event.
 
We apply the same procedure as Otmianowska-Mazur \& Vollmer (2003)
to a similar ram pressure stripping event (Sect.~\ref{sec:model}). We solve
the induction equation: 
\begin{equation}
{\partial\vec{B}/\partial t=\hbox{rot}(\vec{v}\times\vec{B})
 -\hbox{rot}(\eta~\hbox{rot}\vec{B})}
\label{eq:inductioneq}
\end{equation}
where $\vec{B}$ is the magnetic induction, $\vec{v}$ is the large-scale
velocity of the gas, and $\eta$ is the coefficient of a turbulent diffusion,
on a 3D grid ($215\times 215 \times 91$). The cell size is $200$~pc.
The induction equation is solved using a second order Godunov scheme with
second order upstream partial derivatives together with a second order
Runge-Kutta scheme for the time evolution.
This results in less numerical diffusion than that of the ZEUS 3D MHD 
code (Stone \& Norman 1992a,b). We also solved the induction equation using the
ZEUS code with consistent results.
Time-dependent gas-velocity fields are provided by the 3D sticky-particle simulations.
The 3D velocity field obtained from the N-body code has a discrete distribution. The interpolation to
a regular 3D grid was done with the Kriging method (Soida et al. 2006).
We assume the magnetic field to be partially coupled to the gas via the turbulent diffusion process (Elstner et al. 2000) 
assuming the magnetic diffusion coefficient to be $\eta = 5 \times 10^{25}$~cm$^{2}$s$^{-1}$. 
We do not implement any dynamo process. 
The initial magnetic field is purely toroidal. The MHD model does not contain a galactic wind.

The modeled polarized radio continuum emission together with the modeled magnetic field
vectors are shown in the lower panel of Fig.~\ref{fig:n4330_hi_pi}.
The distribution of polarized radio continuum emission has the same extent as it is observed.
This is mainly due to the truncated distribution of relativistic electrons that we assumed.
The distribution is slightly more extended to the northeast. 
As expected, the magnetic field vectors are parallel to the galactic disk.
As in the observations, there is no pronounced asymmetric ridge of polarized emission.
There are two reasons why this is the case: (i) the interaction is slow, i.e. the duration of
ram pressure stripping is shorter than that of NGC~4522 where an asymmetric ridge is observed
(Vollmer et al. 2004). The gas and magnetic field compression is not strong enough to cause an 
asymmetric ridge; (ii) the vector of NGC~4330's 3D velocity vector lies in the plane of the sky.
The interaction thus mainly compresses the dominating azimuthal magnetic field whose vectors are mostly 
parallel to the line of sight. This component of the magnetic field does not contribute to the polarized
radio continuum emission. The absence of an asymmetric ridge of polarized radio continuum emission
is thus well understandable. In addition, a compression by a factor of $2$-$3$ of the azimuthal magnetic field in the
eastern galactic disk would enhance Faraday rotation and thus explain the tilted B vectors in this region.
A direct confirmation of high Faraday rotation measures using polarized emission at 3.6cm and 6cm will represent
a direct proof of gas and magnetic field compression within the disk at these small galactic radii.

\section{Recent massive star formation \label{sec:starformation}}

\begin{figure}
	\resizebox{\hsize}{!}{\includegraphics{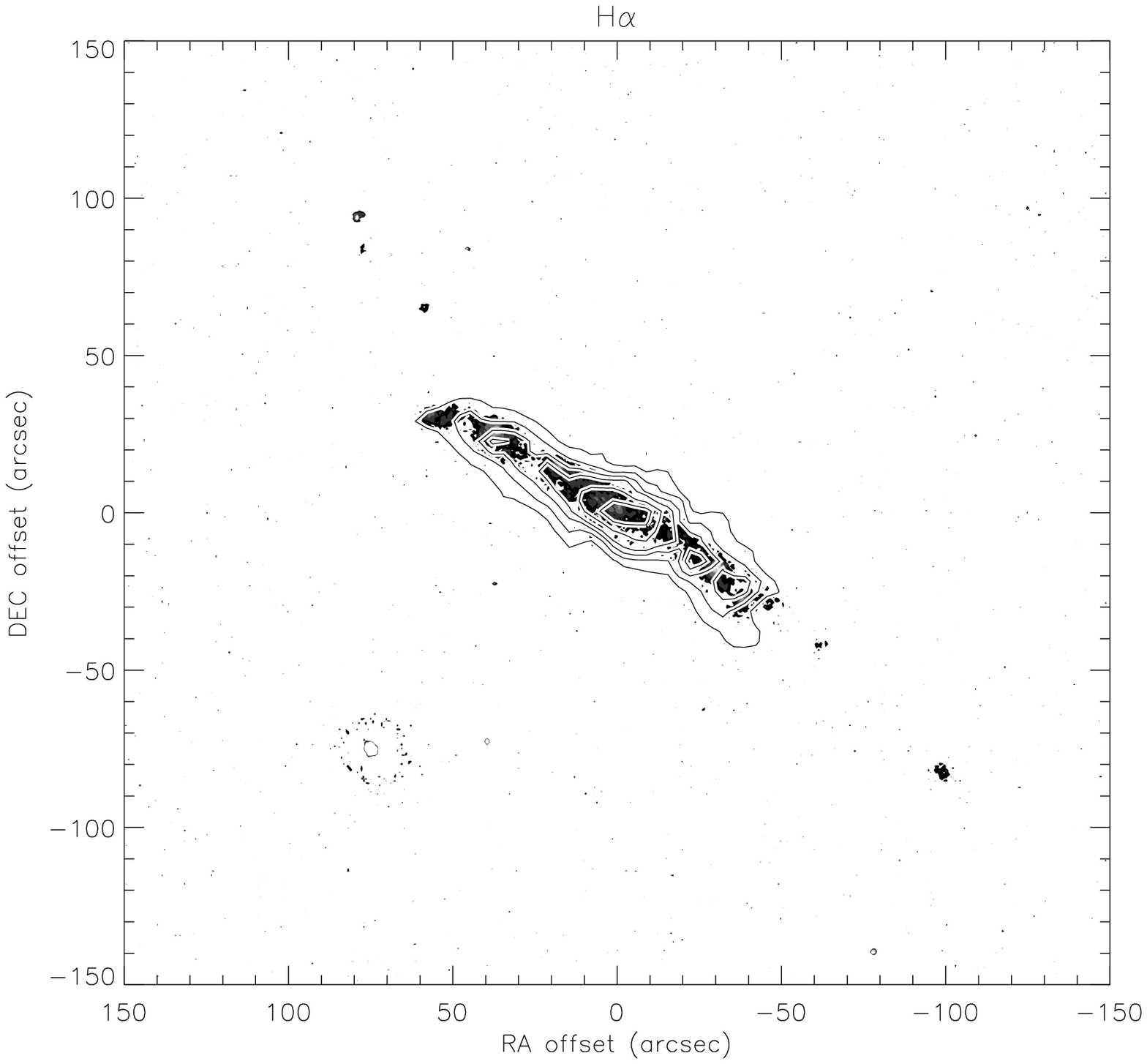}}
        \resizebox{\hsize}{!}{\includegraphics{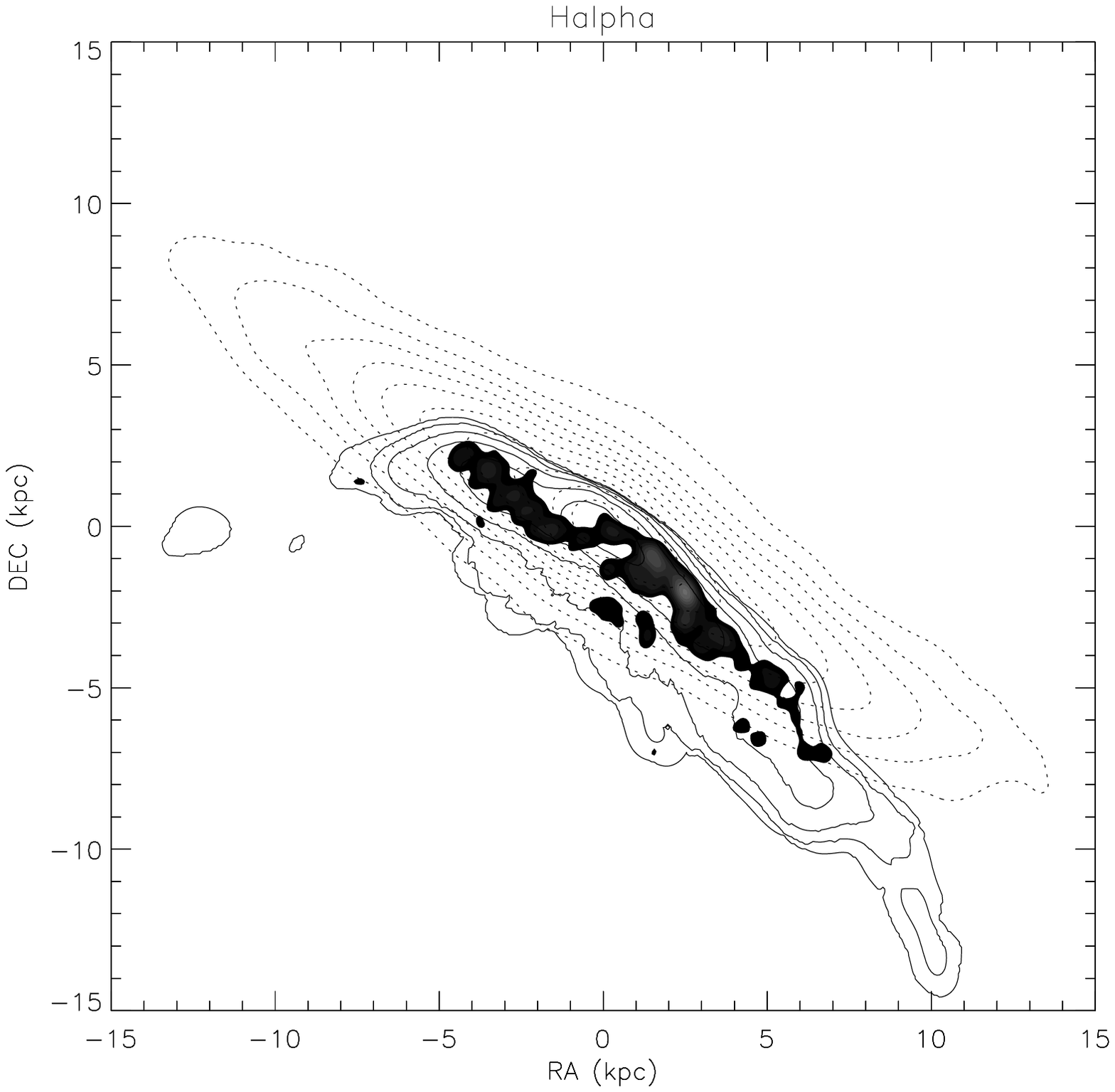}}
	\caption{Recent massive star formation of NGC~4330. Grayscale: H$\alpha$ emission.
          Darker regions are regions with less H$\alpha$ emission. Solid contours:
          molecular gas emission. Upper panel: H$\alpha$ and CO(2--1) observations.
          Lower panel: model. Dotted contours: extinction-free column density of the model stellar disk.
	} \label{fig:n4330_co_halpha}
\end{figure}

Recent massive star formation within the last $\sim 6$~Myr is traced by H$\alpha$ emission.
The upper panel of Fig.~\ref{fig:n4330_co_halpha} shows the H$\alpha$ emission of NGC~4330 
(Abramson et al. 2011) together with the CO(2--1) emission. 
Within the galactic disk the H$\alpha$ emission follows the CO(2--1) emission. The northeastern
upturn is visible in H$\alpha$ and CO(2--1).  On the other hand, the rare extraplanar H$\alpha$ emission
blobs within the extended H{\sc i} tail do not have a CO(2--1) counterpart.
This implies that the bulk of the molecular gas from which the massive stars formed
has been destroyed by stellar winds, SN explosions, and the UV radiation field.
The model H$\alpha$ emission obtained as described in Sec.~\ref{sec:sfr} is shown in the lower
panel of Fig.~\ref{fig:n4330_co_halpha}.
H$\alpha$ emission is mainly found in the high column density gas disk.
Contrary to observations, there is no upturn in the northeast of the galactic in the model H$\alpha$ distribution. 
In addition, some extraplanar H{\sc ii} regions are present south of the model galactic disk.
These do not have observed counterparts.

\section{Star formation within the last 100~Myr \label{sec:sfr100}}

The FUV emission of galactic stellar populations is sensitive to the star formation history during the last $\sim 100$~Myr
whereas the NUV emission decreases on a timescale of $\sim 500$~Myr.
The ratio of FUV to NUV emission depends on the star formation history during the last $500$~Myr and is a
measure of the mean age of the underlying stellar population.
Since the information on the creation time of star particles is kept during the
simulations, it is possible to obtain FUV and NUV emission distributions 
based on STARBURST99 single stellar population fluxes. 
We recall that our star formation recipe based on cloud collisions does only permit an enhanced
star formation rate in stripped gas arms of high volume density (Fig.~\ref{fig:face-on}).
If density inhomogeneities due to turbulent motions in the leeward stripped gas of NGC~4330 lead to
star formation that dominates this region, our model underestimates the star formation of the
stripped extraplanar gas.

\begin{figure}
	\resizebox{\hsize}{!}{\includegraphics{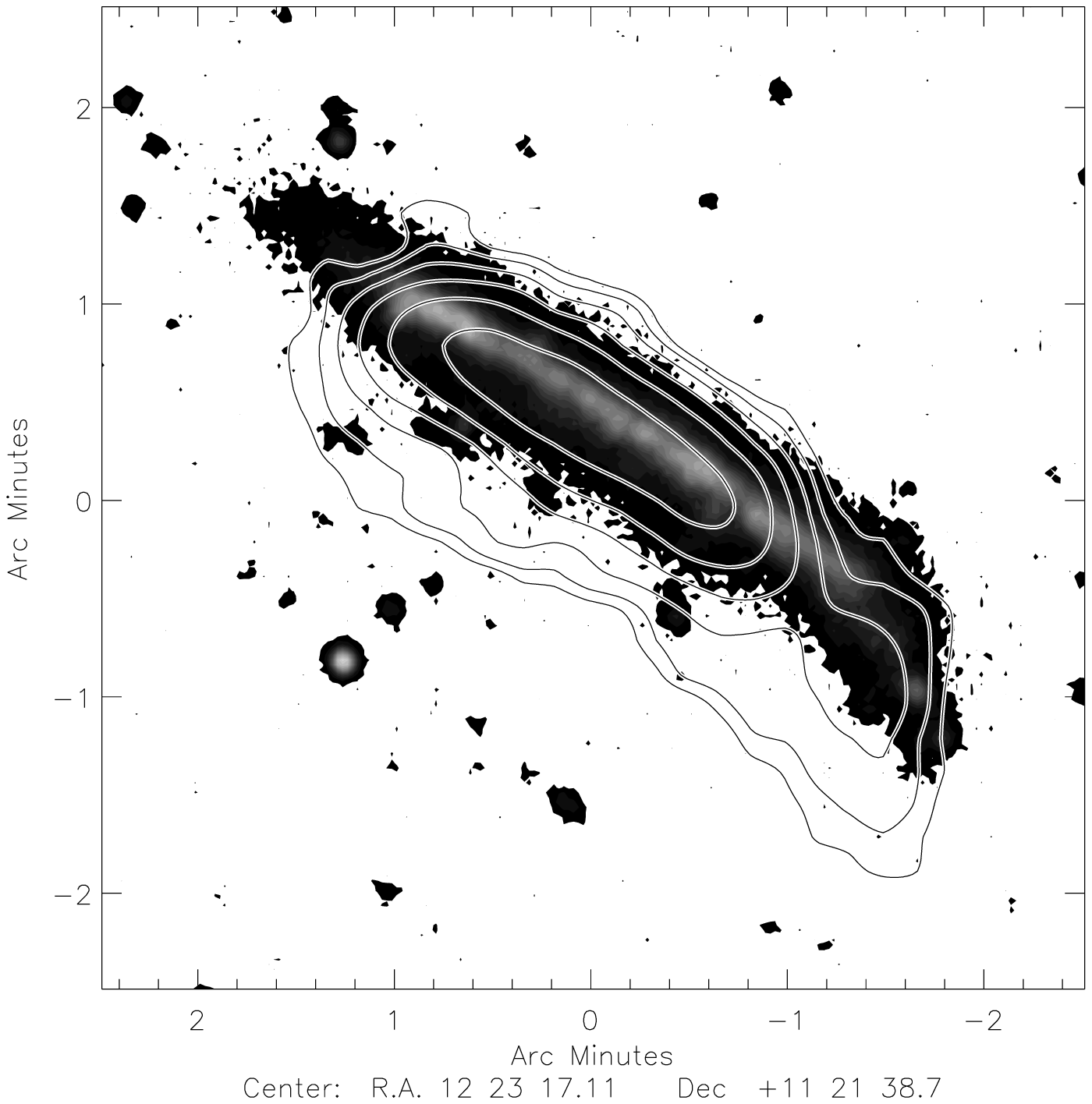}}
	\resizebox{\hsize}{!}{\includegraphics{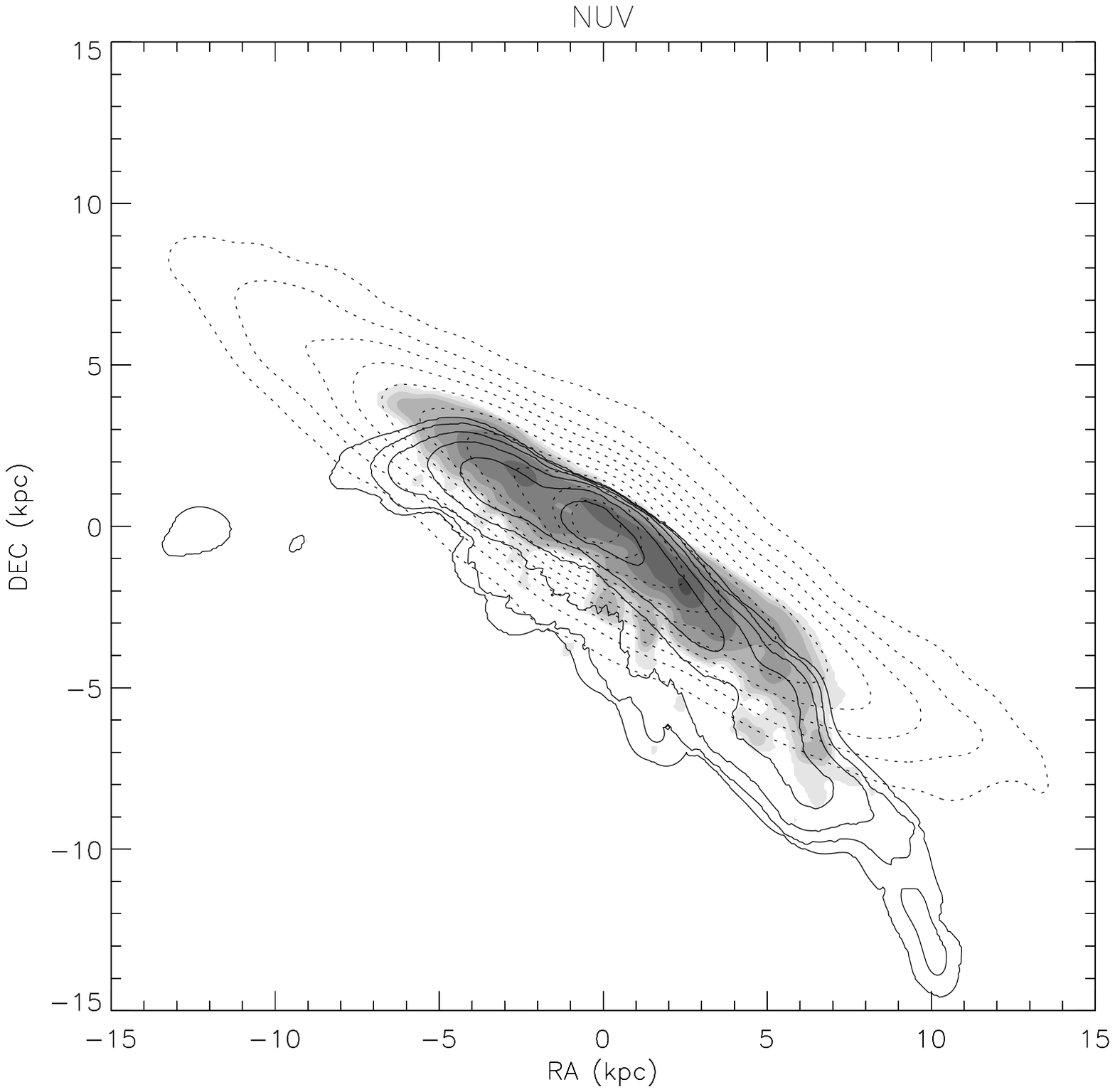}}
	\caption{NGC~4330 -- NUV emission distribution. Upper panel: observations (Abramson et al. 2011). 
	  Contours: H{\sc i} distribution.
	  Lower panel: model (extinction-free). Solid contours: model gas distribution.
	  Dotted contours: extinction-free column density of the model stellar disk.
	} \label{fig:n4330_nuv_hi}
\end{figure}

As shown in Abramson et al. (2011, Fig.~19) the observed NUV 
emission distribution shows a significant offset from the H{\sc i} emission in the extraplanar
southwestern gas tail (upper panel of Fig.~\ref{fig:n4330_nuv_hi}).
The model NUV emission is less extended, but shows qualitatively the same trend 
(lower panel of Fig.~\ref{fig:n4330_nuv_hi}).
Moreover, several isolated extraplanar UV emission blobs are found on the downwind
side to the southeast of the galactic disk (Fig.~21 of Abramson et al. 2011). These blobs are also
present in the model NUV distribution.

\begin{figure}
	\resizebox{\hsize}{!}{\includegraphics{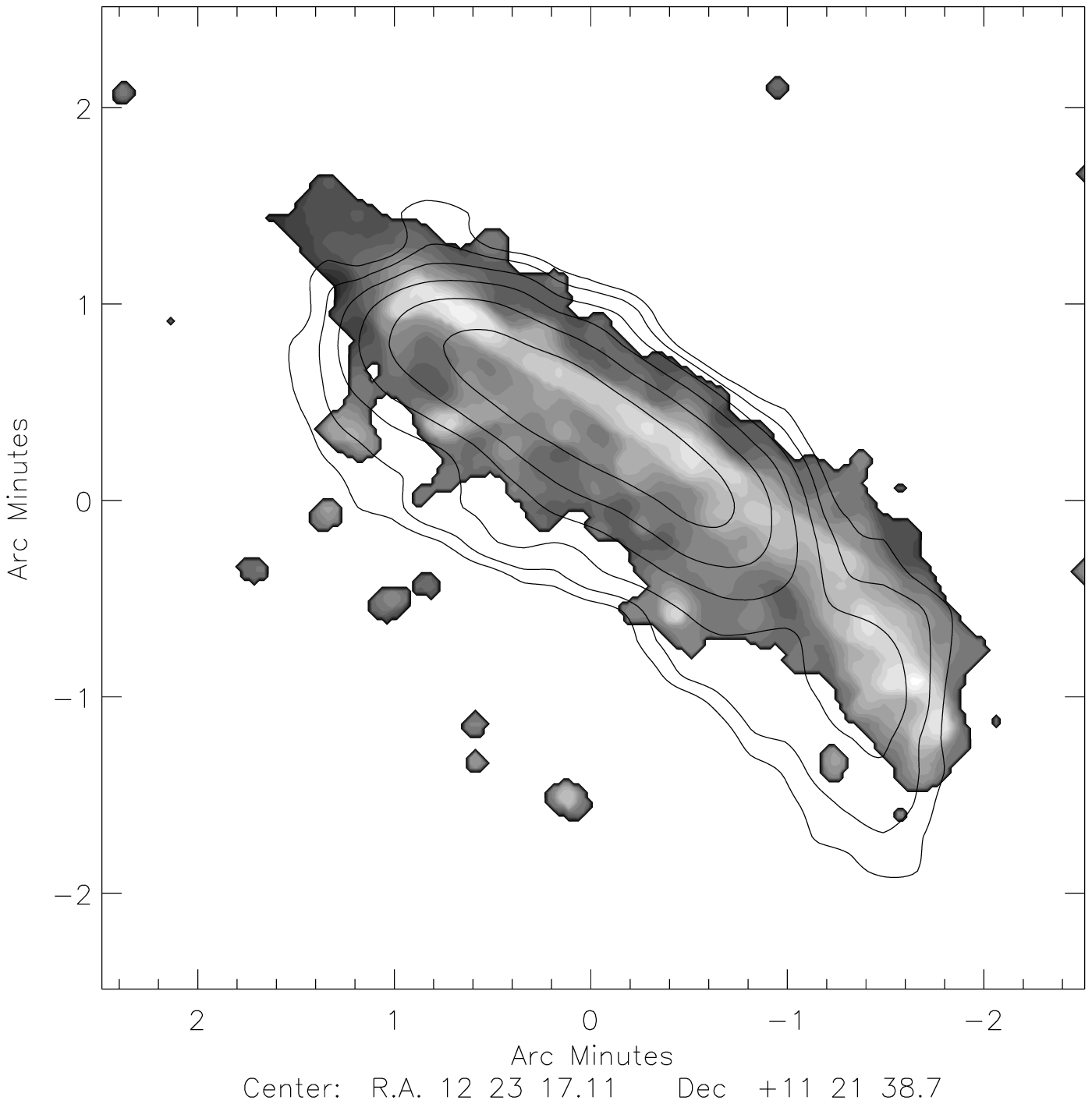}}
	\resizebox{\hsize}{!}{\includegraphics{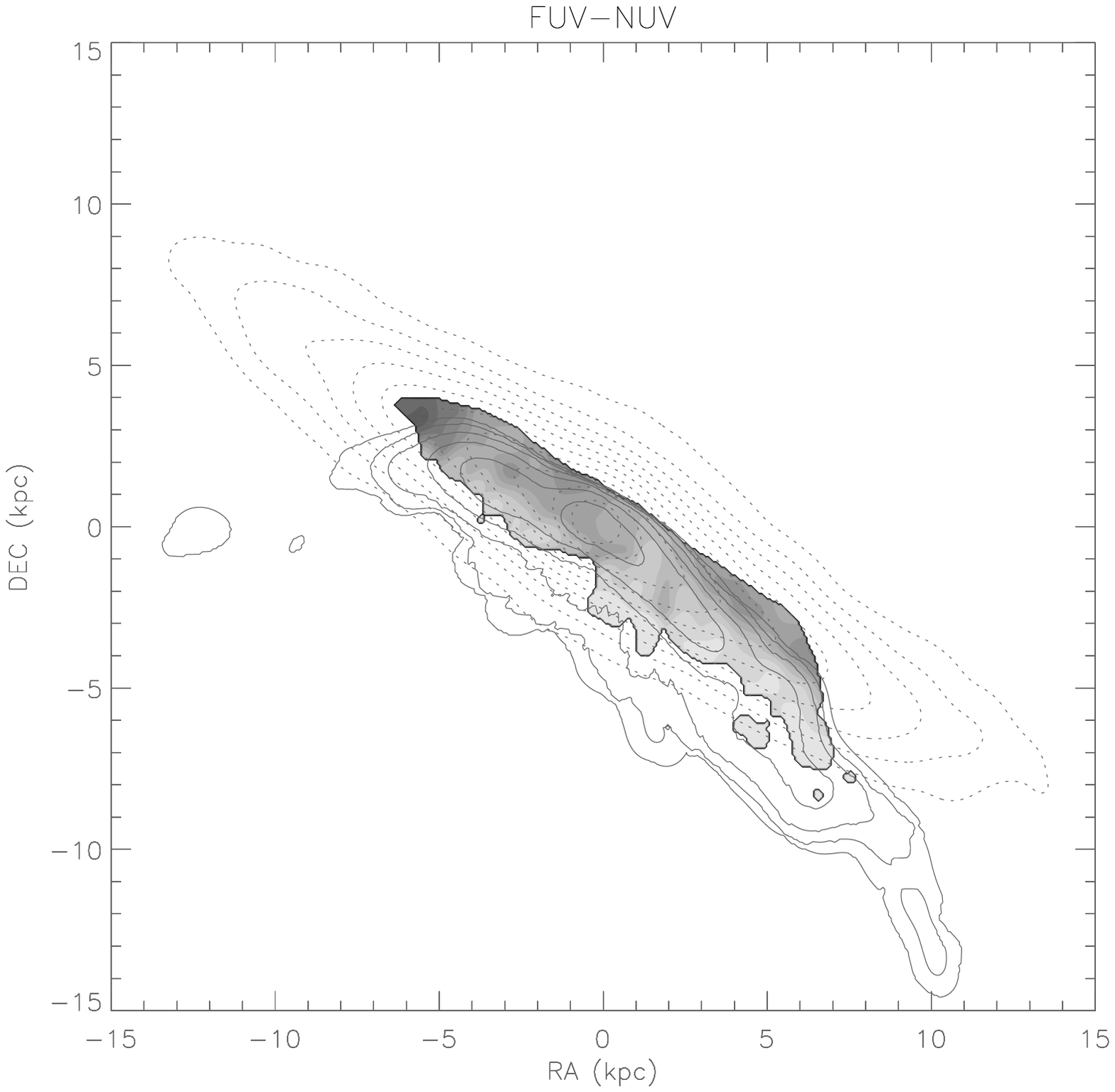}}
	\caption{NGC~4330 -- log(FUV/NUV). Darker areas have a lower log(FUV/NUV). 
	  Upper panel: observations (Abramson et al. 2011). Contours: H{\sc i} distribution.
	  Lower panel: model (extinction-free). Solid contours: model gas distribution.
	  Dotted contours: extinction-free column density of the model stellar disk.
	} \label{fig:n4330_fuv-nuv}
\end{figure}

\begin{figure}
	\resizebox{\hsize}{!}{\includegraphics{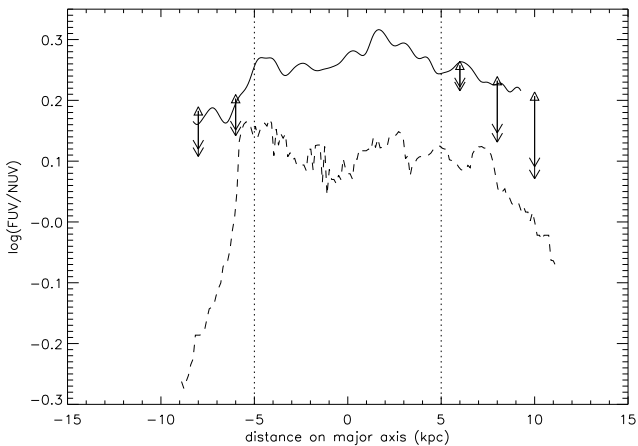}}
	\caption{log(FUV/NUV) along the major axis. 
	  Dashed line: GALEX observations (Abramson et al. 2011).
	  Solid line: model. The vertical dotted lines indicate the gas truncation radius.
	  Triangles: log(FUV/NUV) within 6 radial bins. Arrows: reddening of the
	  UV color due to the inclusion of a constant stellar age distribution
	  during $t=-2$~Gyr and $t=-0.5$~Gyr before the time of interest
	  (c.f. Fig.~\ref{fig:n4330_sfh_model}).
	} \label{fig:n4330_fuv-nuv1}
\end{figure}

The fraction between FUV and NUV emission depends on the age of the underlying stellar populations
modified by extinction. The observed FUV/NUV fraction is high in the stellar disk, the upturn region,
and the extraplanar UV tail (upper panel of Fig.~\ref{fig:n4330_fuv-nuv}). Most of the extraplanar regions
on the downwind side are also FUV bright, i.e. they are young ($\leq 100$~Myr; Abramson et al. 2011).
The extinction-free model FUV/NUV fraction is shown in the lower panel of Fig.~\ref{fig:n4330_fuv-nuv}.
In addition, we show in Fig.~\ref{fig:n4330_fuv-nuv1} cuts along the major axis.
Due to the missing extinction the model FUV/NUV fraction within the gas disk ($R < 5$~kpc) cannot be directly
compared to observations. As observed, the UV emitting regions within the galactic disk
beyond the H{\sc i} truncation radius have a significantly lower FUV/NUV ratio compared to
that of the region within the H{\sc i} distribution. 
The observed variation of $\log({\rm FUV/NUV})$ on the leading side (upturn region) is
much steeper than that of the model. Whereas the observed $\log({\rm FUV/NUV})$ reaches
values $< -0.2$, the model $\log({\rm FUV/NUV})$ never decreases below $0.15$.
On the trailing side (tail region) the difference between the observed and the model
$\log({\rm FUV/NUV})$ is less important than on the leading side, but it is still
significant (up to $\Delta\ \log({\rm FUV/NUV}) \sim 0.2$. 

Our model stellar populations are not older than
$600$~Myr. Since the UV profiles are taken along the major axis within the galactic disk,
old stellar populations of the galactic disk should also be present in the GALEX observations.
If we add a constant stellar age distribution from $-2$~Gyr to $-0.5$~Gyr,
$\log({\rm FUV/NUV})$ decreases by up to $0.1$ in the outer parts of the disk. This is illustrated 
by the arrows in Fig.~\ref{fig:n4330_fuv-nuv1}. Even with this modification, the observed
low values of $\log({\rm FUV/NUV})$ in the upturn region are not attained. 

In the model, all extraplanar regions are young with a $\log({\rm FUV/NUV}) \sim 0.3$, i.e. 
their FUV/NUV ratios are higher than those of the disk. 
Seven out of  9 extraplanar UV-emitting regions found by Abramson et al. (2011)
have similar FUV/NUV fractions ($\log({\rm FUV/NUV}) \geq 0.25$).

We thus conclude that, whereas the observed $\log({\rm FUV/NUV})$ of the tail can be reproduced by 
adding an old stellar population ($2$-$5$~Gyr), the observed $\log({\rm FUV/NUV})$ of the upturn region 
differs largely from the model by $\sim 0.2$~mag.

\section{The overall picture \label{sec:overall}}

From the H{\sc i}, CO, H$\alpha$, UV, and radio continuum observations compared to
the dynamical modeling the following overall picture arises:
NGC~4330 moves approximately northward within the Virgo intracluster medium which is assumed to be static
(see also Vollmer 2009). The line-of-sight velocity component is small compared to the
velocity components in the plane of the sky.
The inclination angle between the disk and orbital plane is $\sim 75^{\circ}$,
i.e. the ram pressure wind blows more face on. This small deviation from a face-on wind is
enough to cause the observed H{\sc i}, CO, H$\alpha$, UV, radio continuum asymmetries between the two extremities 
of the galactic disk. The projected wind angle is between $60^{\circ}$ and $80^{\circ}$.
The current ram pressure acting on NGC~4330 is $p_{\rm rps} \sim 2500$~cm$^{-3}$(km\,s$^{-1}$)$^{2}$.
Based on the orbital segment given in Vollmer (2009) a maximum ram pressure of $\sim 5000$~cm$^{-3}$(km\,s$^{-1}$)$^{2}$
will occur in $\sim 100$~Myr.

On the leading side of the interaction (northeast) an upturn is observed in H$\alpha$, UV (Fig.~15 of Abramson et al. 2011), 
as well as in CO and 6~cm radio continuum emission. We do not find atomic, molecular, or cosmic ray gas beyond 
the upturn region. 
This suggests that the ISM is stripped as a whole entity. It is excluded that gas, which initially has been
left behind, formed rapidly stars, because we do not detect H$\alpha$ or FUV-excess emission sources
beyond the upturn region. The only possible explanation for the absence of dense gas, which might be not affected 
by ram pressure and which might stay behind the less dense gas, is a rapid mixing into the hot intracluster medium.

It is useful to compare the observed properties of NGC~4330 to those of NGC~4522,
another strongly stripped Virgo spiral galaxy.
NGC~4330 has the same inclination angle between the disk and orbital plane and maximum ram pressure as NGC~4522. 
Both H{\sc i} distributions are strongly truncated and show significant extraplanar H{\sc i} emission
(Kenney et al. 2004).
On the other hand, there are three important differences between NGC~4522 and NGC~4330:
(i) whereas NGC~4522 is observed shortly ($\sim 50$~Myr) after peak ram pressure
(Vollmer et al. 2006), NGC~4330 is in a pre-peak
($\sim -100$~Myr) phase, (ii) the width of the ram pressure profile of NGC~4330 is $20$\,\% larger than that of NGC~4522, and
(iii) the dominant component of the 3D velocity vector is along the line-of-sight for NGC~4522, whereas it
is in the plane of the sky for NGC~4330.

Observationally there are two main differences between NGC~4330 and NGC~4522: NGC~4522 shows an asymmetric ridge
of polarized radio continuum emission (Vollmer et al. 2004) and molecular gas beyond the leading edge of the interaction
(Vollmer et al. 2008). Both features are not observed in NGC~4330. 
As already mentioned in Sec.~\ref{sec:mhd}, the absence of an asymmetric ridge of polarized radio continuum emission
can be explained by the projection of NGC~4330: since the galaxy moves predominantly in the plane of the sky,
ram pressure compresses mainly the magnetic field component along the line of sight, which does not
contribute to the polarized radio continuum emission. However, we checked the face-on view of the modeled
polarized radio continuum emission and did not find a pronounced ridge  in the wind direction.
This means that the slowly rising compression of the ISM and its associated magnetic field in NGC~4330 is
still not strong enough to produce a pronounced ridge of polarized radio continuum emission.

Can the absence of CO emission beyond the upturn region in NGC~4330 also be understood by projection effects?
To answer this question we have to compare the interaction timescale, i.e. the time to clear a region
in the galactic disk, the lifetime of a molecular cloud, and the rotation period.
The lifetime of a giant molecular cloud (GMC) is roughly given by its free-fall time.
With a density of $\sim 100$~cm$^{-3}$ the free-fall time is a few Myr.
In NGC~4522 the compression occurs to the east of the galaxy center and rotation transports
the dense molecular clouds that decoupled from the ram pressure wind to the north where they are detected today.
With a rotation velocity of $100$~km\,s$^{-1}$ a molecular cloud at a radius of $3$~kpc
moves by an angle of $\sim 90^{\circ}$ within $50$~Myr.
The interaction timescale is of the order of $\sim 10$~Myr. This implies that the observed
GMCs beyond the gas truncation radius might be longer lived than expected from the free-fall timescale.
A possible mechanism is an enhanced ionization of the molecular gas due to cosmic ray particles
created in the ram pressure compression region.

In NGC~4330 a molecular cloud at the H{\sc i} truncation radius ($\sim 6$~kpc) also needs $\sim 50$~Myr to move by 
$\sim 90^{\circ}$. However, these clouds are then hidden behind the galactic disk.
Thus, only GMCs that recently ($< 20$~Myr) decoupled from the the rest of the ISM can be detected beyond the
gas truncation radius. Since ram pressure is still increasing, more and
more dense clouds should decouple from the wind and stay behind. However, the interaction timescale 
is higher for NGC~4330 (pre-peak with $t_{\rm HW}=100$~Myr) than for NGC~4522 (close to peak with $t_{\rm HW}=80$~Myr).
If the interaction timescale is longer than the lifetime of the GMCs, the ISM is stripped as a whole entity and
the GMCs form stars or are dispersed within the ISM. In both cases, no CO emission is expected beyond the gas truncation radius.
The comparison with NGC~4522 suggests that it needs a higher and/or a more rapidly increasing ram pressure  
so that a significant amount of dense molecular gas ($> 5 \times 10^{6}$~M$_{\odot}$) decouples from the ram pressure wind.

NGC~4330's gas tail consists only of atomic hydrogen and the UV tail is upwind from the H{\sc i} tail.
The case of NGC~4522 (Vollmer et al. 2006) suggested that the extraplanar gas, which is stripped in an arm structure,
forms stars with an efficiency close to that of the galactic disk as long as the large-scale density is high enough.
We checked this possibility for NGC~4330 and found that most of the stars which form the UV tail today are
born in 3 distinct gas arm structures at vertical distances $z < 5$~kpc from the galactic disk within the last $\sim 50$~Myr.
With a free-fall time of a few Myr a collapsing starforming cloud rapidly decouples 
from the ram pressure wind and evolves like a collisionless particle. 
The orbit of the newly created star is thus given by its initial position
and velocity at the moment of decoupling, i.e. they keep the angular momentum of the gas at the time of their
creation. On the other hand, the gas is continuously accelerated by ram pressure gaining constantly angular momentum.
This creates the offset between the UV and H{\sc i} tail.

\begin{figure}
	\resizebox{\hsize}{!}{\includegraphics{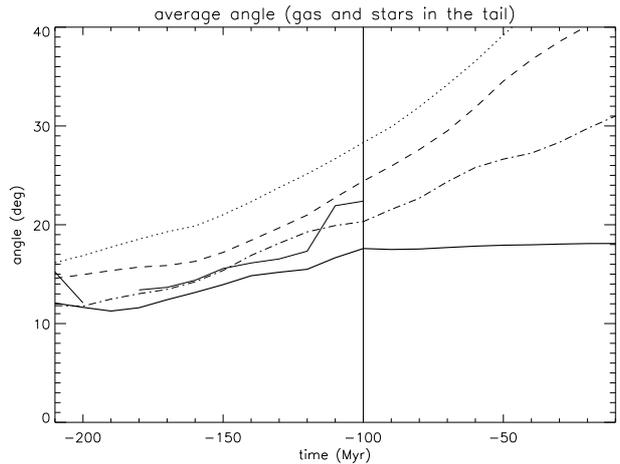}}
	\caption{Evolution of the angle between the angular momentum
          of the disk and that of the (i) gas at three different volume densities (dash-dotted: $n_{0}$,
	  dashed: $0.2 \times n_{0}$, dotted: $n=0.04 \times n_{0}$), (ii) newly born tail stars younger than $10$~Myr
          (solid line), and (iii) all young tail stars with ages smaller than $200$~Myr (thick solid line). 
	} \label{fig:angle_deviation}
\end{figure}

It is possible to verify this mechanism with our dynamical simulation. For this we calculated the deviation angle of
(i) the angular momentum of the newly created tail stars and (ii) the extraplanar gas tail of different densities 
from the angular momentum of the disk (Fig.~\ref{fig:angle_deviation}). 
The tail region for the stars is defined in the following way: $3$~kpc~$\leq$~$RA$~$\leq$~$10$~kpc and 
$-10$~kpc~$\leq$~$DEC$~$\leq$~$-5$~kpc on Fig.~\ref{fig:n4330_fuv-nuv}. 
The tail region for the gas is defined as the region with a vertical distance $> 1.2$~kpc from the
galactic disk.
The small-scale variation in the evolution of the deviation angle of the young stars is due to individual
stars entering and leaving the tail region. The offset between the deviation angle 
of the young stars with ages smaller than $200$~Myr and that of the gas increases with decreasing gas density. 
Low density gas is thus pushed further away from the disk than the newly created stars which form in
high density gas. 
The gas tail is thus more and more pushed away from the UV tail which consists of stars that are no longer affected by
ram pressure.

Whereas the H{\sc i}, CO, H$\alpha$, and dust extinction observations suggest a recent stripping of the
outer galactic disk ($\sim 100$~Myr), the FUV-NUV and NUV-r colors are in favor of a much older stripping ($\geq 200$~Myr)
of these regions (Abramson et al. 2011 and Fig.~\ref{fig:n4330_fuv-nuv1}). 
As discussed by Abramson et al. (2011), the FUV-NUV and NUV-r colors cannot be explained by simple quenching models
of a previously constant star formation history. 
The reconstruction of the stellar age distribution from deep optical spectra (Crowl \& Kenney 2008, Pappalardo et al. 2010) 
might solve the problem.

\section{The stellar age distribution of the outer disk \label{sec:sfrhist}}

\begin{figure}
	\resizebox{\hsize}{!}{\includegraphics{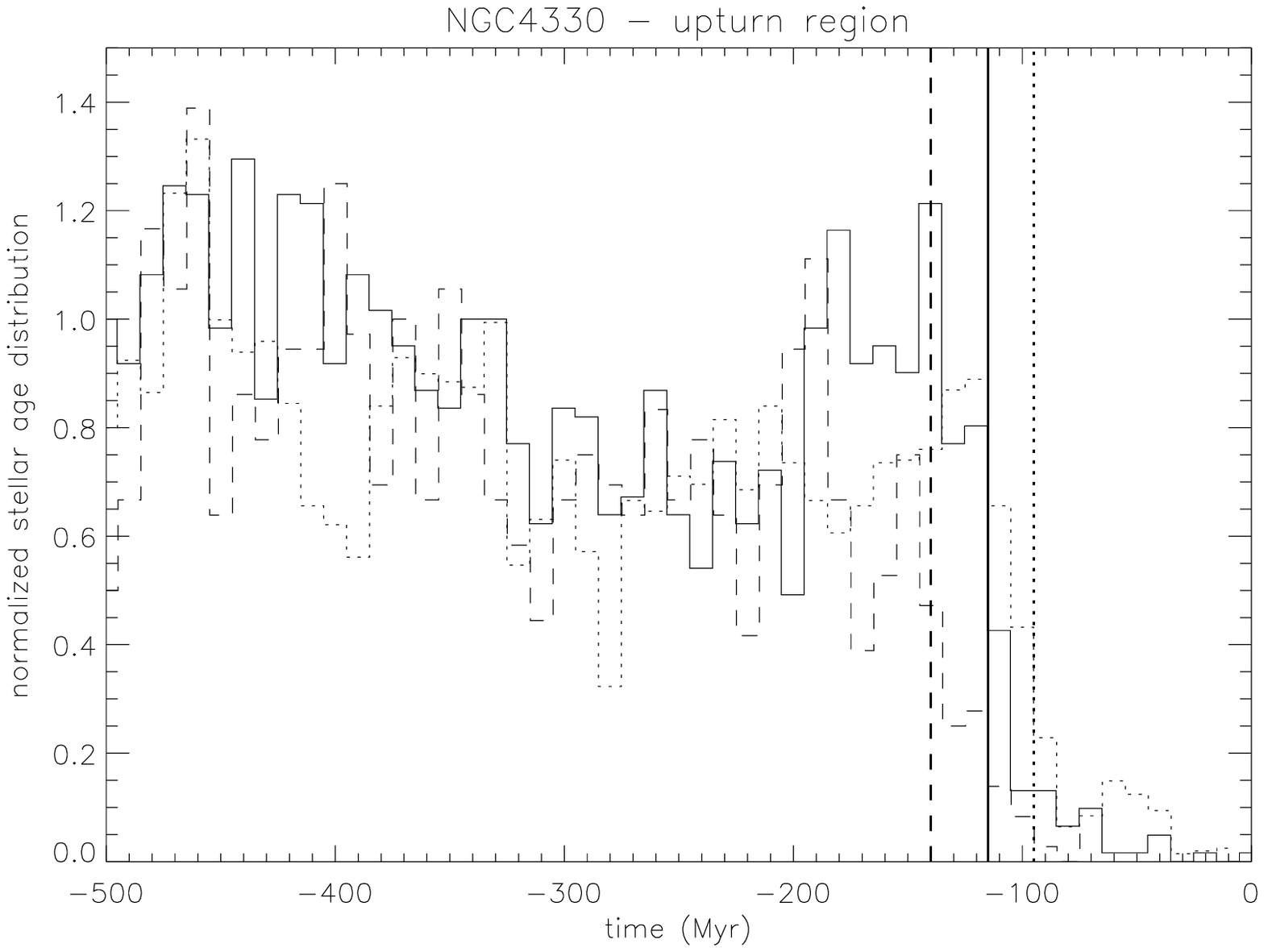}}
	\resizebox{\hsize}{!}{\includegraphics{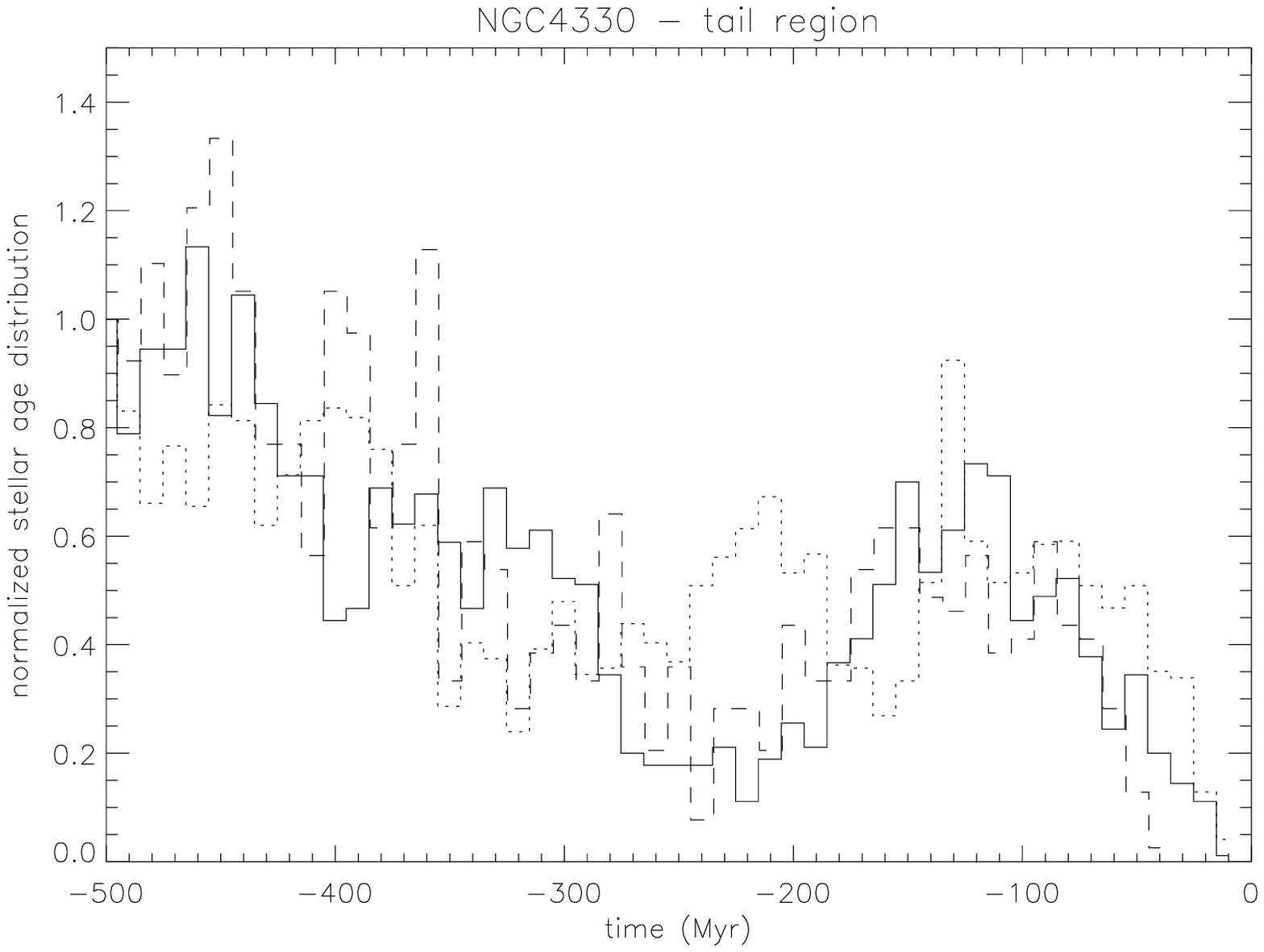}}
	\caption{Model stellar age distribution within radial bins outside the gas truncation radius.
	  Dotted line: $5$~kpc$ < R < 7$~kpc. Solid line: $7$~kpc$ < R < 9$~kpc. Dashed line: $9$~kpc$ < R < 11$~kpc.
	  Upper panel: upturn region.
	  The vertical lines correspond to the quenching time for which the age distribution has dropped by
	  a factor 2 compared to its initial value.
	  Lower panel: gas tail region.
	} \label{fig:n4330_sfh_model}
\end{figure}

As discussed in Sect.~\ref{sec:sfr} stars are created during cloud-cloud collisions in the dynamical model.
Each newly created star has the information on its creation time during the simulation which begins
$600$~Myr before the ram pressure maximum. In the following we define $t=0$~Myr as the time of observations,
i.e. $100$~Myr before peak ram pressure. Stellar age distributions are determined within the galactic
disk ($|z| < 1$~kpc) within 3 radial bins outside the gas truncation radius: (i) $5$~kpc$ < R < 7$~kpc,
(ii) $7$~kpc$ < R < 9$~kpc, and (iii) $9$~kpc$ < R < 11$~kpc. This corresponds to the age distributions of the
newly created stellar particles within the radial bins (Fig.~\ref{fig:n4330_sfh_model}).
Following Pappalardo et al. (2010) we define the quenching time as the time for which the age distribution has dropped by
a factor 2 compared to its initial value. 

In the upturn region there is a clear gradient of quenching times: radial bin (i) $-95$~Myr, (ii) $-130$~Myr,
(iii) $-160$~Myr. On the other hand, the quenching times of the outer two radial bins in the tail
region are $\sim -350$~Myr. After this drop the stellar age distribution stays approximately constant
until $-50$~Myr when it begins to decrease again. The stellar age distribution of the inner radial bin is that of
a normally starforming disk with passing spiral arms which are responsible for the different peaks.
In this disk area the model gas disk has still a high surface density contrary to observations (Fig.~\ref{fig:ngc4330.pvd}).

\begin{figure}
	\resizebox{7cm}{!}{\includegraphics{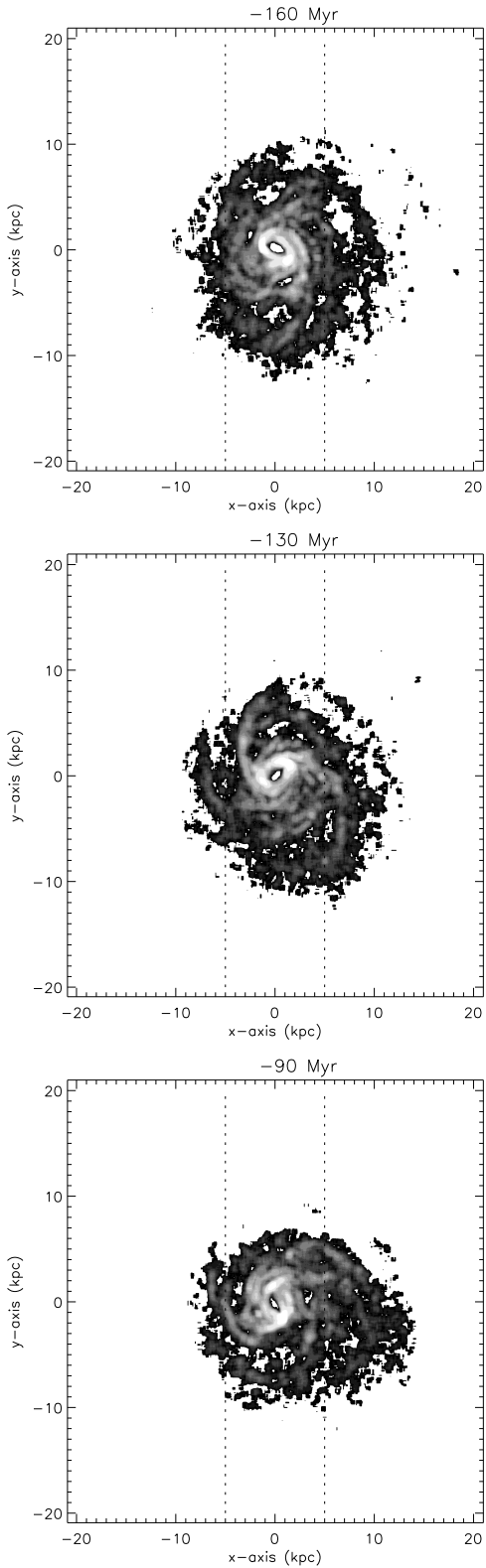}}
	\caption{Model gas surface density distribution within the galactic disk
	  ($|z| < 1$~kpc) at the 3 quenching times of the upper panel of
	  Fig.~\ref{fig:n4330_sfh_model}. Darker regions correspond to lower surface densities.
	  The timestep of the snapshots are marked on top of each panel, $t=0$ corresponds to maximum ram pressure.
	  The ram pressure wind blows from the left side.
	  The vertical lines represent a truncation radius of $5$~kpc on the leading and
	  trailing side of the interaction at $t=0$.
	} \label{fig:gasdisksnapshots}
\end{figure}

It is not straight-forward to link the stellar age distribution to the evolution of the gas surface density, because
the structure of the gas disk is rapidly evolving during the ram pressure stripping event.
Existing spiral arms are reshaped and sometimes additional ones are created. Stars mainly form in the dense gas
of these spiral arms. To illustrate the situation, we show in Fig.~\ref{fig:gasdisksnapshots} the gas surface 
density distributions of the 3 quenching times of the upper panel of Fig.~\ref{fig:n4330_sfh_model}.
At each timestep the gas disk has a complex asymmetric spiral structure with 3 main arms.
At $t=-90$~Myr the spiral arm at the leading side is at a distance of $5$~kpc.
It is not obvious to define a gas truncation radius on the leading (left) side of the interaction to compare it
to the quenching times. 

Nevertheless, we determined the radius at which the gas surface density within the disk ($|z| < 1$~kpc)
drops to a given value $\Sigma_0$ at the
leading (left) and trailing (right) side of the interaction. $\Sigma_0$ is chosen in a way to give the
same truncation radius at both sides today. 
\begin{figure}
	\resizebox{\hsize}{!}{\includegraphics{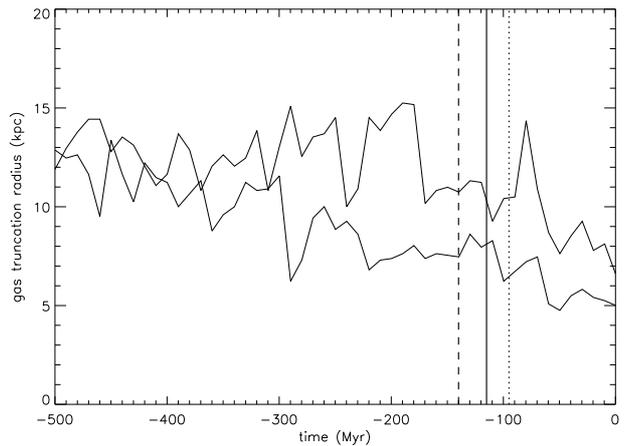}}
	\caption{Model gas truncation radius at the leading (lower graph) and trailing (upper graph) side
	  of the interaction. The vertical line show the quenching ages of the upper panel of
	  Fig.~\ref{fig:n4330_sfh_model}.
	} \label{fig:gastruncation}
\end{figure}
Due to the non-zero inclination angle between the disk and the ram pressure wind of $25^{\circ}$, the gas
is pushed within the galactic disk towards the right. Thus, the trailing side of the gas disk is more extended than
the leading side for most of the considered time interval. At the same time, the gas is pushed
vertically out of the disk plane. This symmetrizes the gas distribution within the galactic disk
($|z| < 1$~kpc). The variation of the truncation radius at small timescales is due to the passage of
spiral arms. Between $-150$~Myr and $-80$~Myr the truncation radius of the leading side decreases more
rapidly than that of the trailing side. This trend is, however, much less clear than the signature of ram pressure 
in the evolution of the quenching time.
We thus conclude that it is necessary to include star formation over the last $\sim 0.5$~Gyr to
make a meaningful comparison between the observed and model quenching times.

\section{Conclusions \label{sec:conclusions}}

In the series of detailed examinations of ram pressure stripped Virgo spiral galaxies, NGC~4330
is the second case after NGC~4501 where the galaxy is still approaching the cluster
center and  ram pressure is still increasing. The edge-on projection makes
the identification of extraplanar gas and star formation possible. Abramson et al. (2011)
showed that the H$\alpha$ and dust extinction distributions show an upturn at the leading side
of the interaction (their Figs.~7 and 15). On the opposite side H{\sc i} and UV tails are detected with a significant offset
between the two tails. The UV colors show a radial age gradient of the youngest stellar populations
in the gas-free part of the upturn region. On the tail side no such gradient has been found.

In this article we present new CO(2-1) and 6~cm radio continuum emission observations of NGC~4330.
The available multiwavelength observations (UV, H$\alpha$, H{\sc i}, CO, 6~cm total power radio continuum,
6~cm polarized radio continuum) are compared to a dynamical model including ram pressure stripping
and, for the first time, star formation. 
The distribution of molecular gas in the galactic disk is more extended towards the upturn region.
The upturn itself is visible in CO and radio continuum emission. No CO nor radio continuum emission 
is detected in the gas tail.  As the H{\sc i} emission, the
6~cm total power radio continuum emission is less extended along the minor axis on the windward side.
There is a second radio continuum upturn region on the tail side. This second upturn is only
visible in radio continuum emission.
The polarized radio continuum emission is relatively symmetric and, in contrast to other
Virgo spirals affected by ram pressure, does not show an asymmetric ridge in the outer disk.

The best fit model has been chosen from a series of simulations with different (i) inclination angles
between the ram pressure wind and the disk plane, (ii) values of the maximum ram pressure, and (iii)
durations of the ram pressure stripping event. The best fit model is consistent with the
galaxy's projected position and radial velocity in the Virgo cluster and the model gas distribution
and velocity field reproduce the H{\sc i} observations. NGC~4330 experiences a ram pressure of $p_{\rm rps}=
2500$~cm$^{-3}$(km\,s$^{-1}$)$^{2}$ and the angle between the ram pressure wind and the galactic
plane is $75^{\circ}$. Based on the orbital segment given in Vollmer (2009) a maximum ram pressure of 
$\sim 5000$~cm$^{-3}$(km\,s$^{-1}$)$^{2}$ will occur in $\sim 100$~Myr.

UV, H$\alpha$, CO, and H{\sc i} maps are derived from the best fit model snapshot.
From the comparison between the dynamical model and observations we conclude that
\begin{enumerate}
\item
the model reproduces qualitatively the observed atomic and moleceular gas distribution.
The morphology of the model gas tail is different for simulations with and without gas shadowing.
The models are unable to reproduce simultaneously the observed discontinuity and angle of the gas tail.
\item
in the framework of the model it is not possible to explain the much stronger bending of the radio continuum tail
with respect to the gas tail. More efficient stripping of the cosmic ray gas or enhanced
diffusion of cosmic ray electrons might be at work, 
\item
the absence of an asymmetric ridge of polarized radio continuum emission is due to
a slow compression of the ISM and the particular projection of NGC~4330 with a small
line-of-sight component of the galaxy's 3D velocity vector,
\item
the spatial offset between the H{\sc i} and UV tail is well reproduced. Since collapsing and
starforming gas clouds decouple from the ram pressure wind, the UV-emitting young stars
have the angular momentum of the gas at the time of their creation. On the other hand,
the gas is constantly pushed by ram pressure and is thus constantly changing angular momentum, 
\item
the UV color profile along the major axis is reproduced qualitatively by the model
with a smaller FUV/NUV ratio in the northeastern upturn region compared to that of the tail region.
However, the observed very low FUV/NUV ratio in the upturn region is not reproduced by the model.
Whereas the H{\sc i}, CO, H$\alpha$, dust extinction observations,
and the dynamical model suggest a more recent gas stripping of the outer galactic disk
($\sim 100$~Myr), the UV color favors an older stripping ($\geq 200$~Myr) of this region (Abramson et al. 2011).
To investigate this discrepancy, we provide stellar age distributions within 3 radial bins 
in the galactic disk ($R > 5$~kpc).
The star formation quenching ages based on these stellar age distributions can be
verified by the stellar population synthesis inversion of deep optical spectra
(Crowl \& Kenney 2008, Pappalardo et al. 2010).
\item
due to a complex evolving gas distribution, it is necessary to include star formation over the 
last $\sim 0.5$~Gyr in the model to make a meaningful comparison between the observed and model 
star formation quenching times.
\end{enumerate}

We are now at the point where we can study the reaction (phase change, star formation) 
of the multiphase ISM (molecular, atomic, ionized) to ram pressure. It seems that the ISM
is stripped as a whole entity. Only a tiny fraction of dense clouds can be left
behind (see also Vollmer et al. 2006). Extraplanar star formation proceeds in dense gas arms
pushed by ram pressure. The collapsing clouds decouple rapidly from the ram pressure wind
and the young, UV-emitting stars stay behind the gas. More high resolution multiwavelength
studies of ram pressure stripped galaxies are necessary to confirm and refine these
hypotheses.

\begin{acknowledgements}
B.V. would like to thank the MPIfR (P. Reich) for computational support. 
This work has been supported by the Polish Ministry of Science and Higher Education
grants 92/N-ASTROSIM/2008/0 and 3033/B/H03/2008/35.
This work also has been supported by the National Research Foundation of Korea grant 2011-8-0993,  
Yonsei research grant 2010-1-0200 and 2011-1-0096. Support was also provided by the National Research 
Foundation of Korea to the Center for Galaxy Evolution Research.
\end{acknowledgements}

\appendix

\section{The model star formation law}

\begin{figure}
	\resizebox{8cm}{!}{\includegraphics{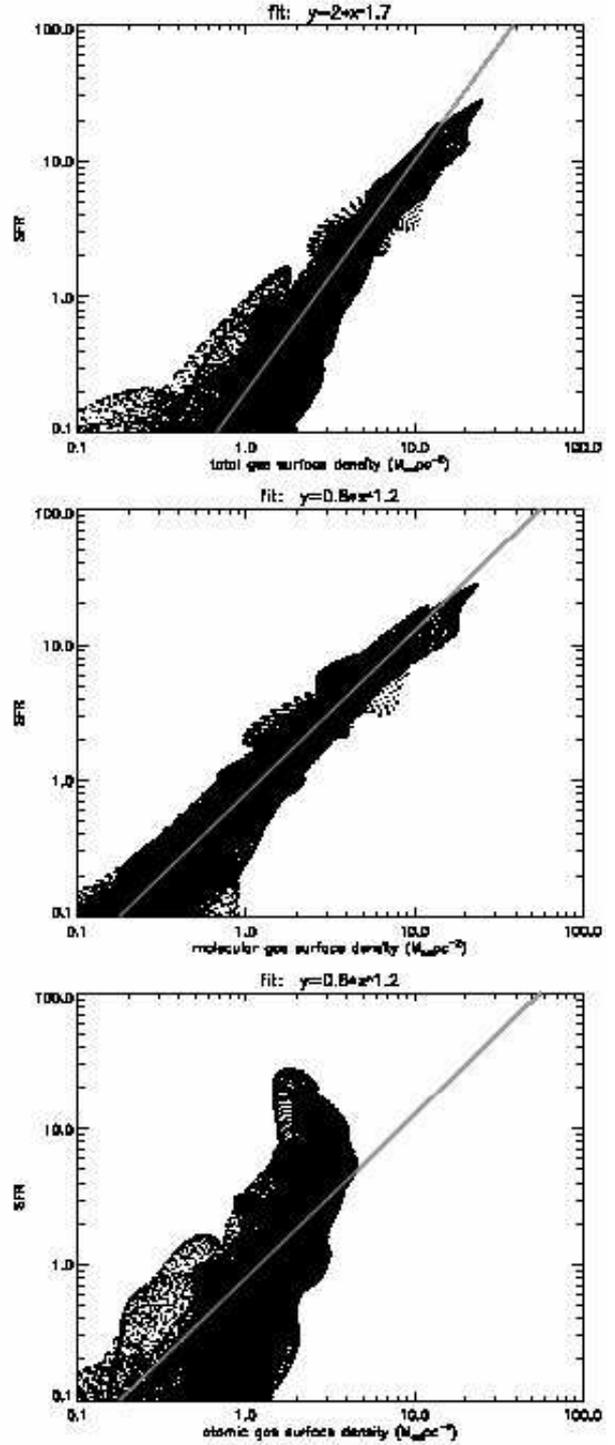}}
	\caption{The model star formation law. Upper panel: as a function of the total gas surface density.
	  Middle panel: as a function of the molecular gas surface density. Lower panel: as a function
	  of the atomic gas surface density. The solid lines show power law fits with exponents 1.7 (total),
	  and 1.2 (molecular). The solid line in the lower panel is there to guide the eye.
	  Star formation is assumed to be proportional to the unobscured model UV emission (in arbitrary units).
	  The molecular gas surface density is calculated by assuming $\Sigma_{\rm mol}/\Sigma_{\rm tot} \propto
	  1.4 \sqrt{\rho}$, where $\rho$ is the total gas volume density. The atomic gas surface density is
	  calculated by assuming $\Sigma_{\rm HI}/\Sigma_{\rm tot} \propto 1 - 1.4 \sqrt{\rho}$.
	} \label{fig:schmidtlaw}
\end{figure}

\end{document}